\documentclass[oneside,english]{amsart}
\usepackage[T1]{fontenc}
\usepackage[latin9]{inputenc}
\usepackage{geometry}
\usepackage{float}
\usepackage{amsbsy}
\usepackage{amsthm}
\usepackage{amssymb}
\usepackage{graphicx}
\usepackage{setspace}
\usepackage[authoryear]{natbib}

\makeatletter

\newcommand{\lyxdot}{.}

\numberwithin{equation}{section}
\numberwithin{figure}{section}

\usepackage{rotating}

\newcommand{\specialcell}[2][c]{%
  \begin{tabular}[#1]{@{}c@{}}#2\end{tabular}}

\makeatother

\usepackage{babel}

\begin{document}
\setstretch{1.5} 
\title{Variance Components Genetic Association Test for Zero-inflated Count
Outcomes}

\maketitle

\smallskip{}

\begin{center}
\textbf{Matthew Goodman}\\
Department of Biostatistics, Harvard T.H. Chan School of Public Health,
Boston, MA 02115, USA
\par\end{center}

\begin{center}
\textbf{Lori Chibnik }\\
Department of Epidemiology, Harvard T.H. Chan School of Public Health,
Boston, MA 02115, USA
\par\end{center}

\begin{center}
\textbf{Tianxi Cai}\\
Department of Biostatistics, Harvard T.H. Chan School of Public Health,
Boston, MA 02115, USA
\par\end{center}

\smallskip{}

\begin{center}
\textit{Correspondence:} Matthew Goodman\\

Department of Biostatistics

Harvard T.H. Chan School of Public Health\\

Boston, MA 02115, USA\\

\textit{Phone:} 617.432.1056\\

\textit{Email:} matthewgoodman@fas.harvard.edu\pagebreak{}
\par\end{center}

\pagebreak{}

\begin{abstract}
Commonly in biomedical research, studies collect data in which an
outcome measure contains informative excess zeros; for example when
observing the burden of neuritic plaques in brain pathology studies,
those who show none contribute to our understanding of neurodegenerative
disease. The outcome may be characterized by a mixture distribution
with one component being the `structural zero' and the other component
being a Poisson distribution. We propose a novel variance components
score test of genetic association between a set of genetic markers
and a zero-inflated count outcome from a mixture distribution. This
test shares advantageous properties with SNP-set tests which have
been previously devised for standard continuous or binary outcomes,
such as the Sequence Kernel Association Test (SKAT). In particular,
our method has superior statistical power compared to competing methods,
especially when there is correlation within the group of markers,
and when the SNPs are associated with both the mixing proportion and
the rate of the Poisson distribution. We apply the method to Alzheimer's
data from the Rush University Religious Orders Study and Memory and
Aging Project (ROSMAP), where as proof of principle we find highly
significant associations with the APOE gene, in both the `structural
zero' and `count' parameters, when applied to a zero-inflated neuritic
plaques count outcome.\medskip{}

\end{abstract}


\subsection*{Acknowledgements:}

This research was funded in part by NIH grant T32NS048005. The content
is solely the responsibility of the authors and does not necessarily
represent the official views of the National Institutes of Health.
The authors report no conflict of interests.

\pagebreak{}

\section{Introduction}

It is common to encounter data that would be appropriate to model
with a so-called zero-inflated distribution, where some proportion
of subjects obtain a zero outcome while the remainder obtain a positive
outcome from some standard distribution. In analyzing such data in
order to discover associations with genetic markers, it is important
to use statistical tests that are adapted to this setting, both in
order to appropriately control type I error and maximize power. This
can present a challenge for naive approaches. For example, dichotomizing
the data or disregarding the zeros is obviously problematic because
each of these approaches results in data loss, ignoring information
that can contribute to our understanding. Another simple approach
uses the available zero-inflated data, but fails to account for zero-inflation.
Namely, one could perform a multivariate Wald test using the Huber-White
robust variance estimator within the usual Poisson regression. However,
in our simulations we find that this last approach can fail to preserve
type I error. In this paper we propose an association testing procedure,
modeling the outcome via zero-inflated Poisson (ZIP) regression \citep{lambert1992},
a simple yet flexible model that can be used to infer about the underlying
mixture of two subgroups: (i) the `structual zero' group representing
a `healthy' or unaffected population; and (ii) the `susceptible group'
with varying degrees of severity captured by a Poisson distribution. 

We aim to develop testing procedures that can effectively assess the
overall association between a set of genetic markers and a zero-inflated
count outcome. SNP-set analyses have been advocated as having several
advantages over standard single-SNP analyses including better reproducibility,
power and interpretability \citep{liu2007,liu2008,wu2010}. For example,
with gene-level SNP sets, several SNPs may affect transcription levels
of a given protein. Individual SNP effects may combine additively
or may involve more complex interactions. Marginal testing of individual
SNPs may miss important signals, first due to low power from inability
to combine weak effects across multiple SNPs, and second due to poor
model fit from inability to model interactions or other complex effects.
Marginal testing also suffers from additional multiple-comparisons.
Variance components score tests for semi-parametric kernel regression
been previously shown to outperform standard multivariate tests with
$p$ degrees of freedom in these of these respects, within the continuous,
binary, and time-to-event outcome settings \citep{wu2010,wu2011,cai2011,shen2016}.

By modeling the zero-inflated count outcome via ZIP or zero-inflated
negative binomial (ZINB) \citep{greene1994}, one may form Wald tests
to assess the overall association between a SNP set and the outcome.
However, such Wald tests have limited power when the SNPs are in high
linkage disequilibrium (LD) with each other. Furthermore, these simple
Wald test procedures cannot easily accommodate non-linear effects
and are sensitive to model mis-specification. To overcome these challenges,
we propose in this paper a variance component test via the ZIP framework
to detect signals from both the genetic effect on the mixing proportion
and on the Poisson rate. (Since it is not our focus here, we refer
the reader to references in the literature, e.g. \citep{ridout2001score}
where various tests have been proposed for detecting when a data set
is zero-inflated, or if it is zero-inflated, whether there is overdispersion
in the Poisson model.)

The rest of the article is organized as follows. In Section 2, we
introduce the ZIP kernel-machine setting and we present the score
test and procedures for approximating the null distribution of the
proposed test. Simulation results are presented in Section 3 and the
proposed procedures are illustrated by assessing the association of
a zero-inflated neuritic plaques count phenotype with gene-level SNP-sets
in the ROSMAP cohort. We conclude with a discussion.

\section{Methods}

\subsection{Description of Data}

Suppose we have genetic data for $n$ subjects with a set of $p$
SNPs and $q$ covariates. Let $Y_{i}$, $\mathbf{{G}}{}_{i}=(G_{i1},G_{i2},..,G_{ip})$,
$\mathbf{{X}}{}_{i}=(X_{i1},X_{i2},..,X_{iq})$, indicate respectively
the the $i^{th}$ subject's outcome, genotypes, and covariates. We
assume that $\mathbb{{D}}=\{{\mathbf{D}}_{i}=(Y_{i},\mathbf{{G}}_{i}^{\top},\mathbf{X}_{i}^{\top})^{\top},i=1,...,n\}$
are independent and identically distributed random vectors. Covariates
might include age, gender, and, to control population stratification,
top principal components of the genetic covariance matrix. For our
examples, genotypes $G_{ij}=0$, $1$, or $2$ represent the number
of minor alleles at a given locus under the assumption of an additive
genetic model. However, one can recode genotypes if dominant or recessive
models are desirable. The count outcome phenotype $Y_{i}$ is assumed
to follow a ZIP model consisting of two components: 1) structural
zeros with probability $1-\pi_{i}$, , and 2) a Poisson($\lambda_{i}$)
distribution with probability $\pi_{i}$. (Note that the parameterization
of $\pi_{i}$ is nonstandard, but is here chosen so the direction
of effect in the $\pi_{i}$ regression and the $\lambda_{i}$ regression
agree qualitatively in sign.) 

Our goal is to devise a testing procedure to assess whether $\mathbf{{G}}{}_{i}$
plays a role in either $\lambda_{i}$ or $\pi_{i}$. To this end,
we note that the ZIP likelihood takes the form

\begin{equation}
\prod_{i=1}^{n}\mathcal{L}(\pi_{i},\lambda_{i}|\mathbf{D}_{i})=\prod_{i=1}^{n}\{(1-\pi_{i})I(Y_{i}=0)+\pi_{i}(e^{-\lambda_{i}}\lambda_{i}^{Y_{i}}/Y_{i}!)\}\label{eq:model}
\end{equation}
We further parameterize the ZIP model in semi-parametric fashion,
where we assume the effect of $\mathbf{{G}}_{i}$ on $Y_{i}$ is fully
captured by the functions $h_{\pi}(\mathbf{{G}}_{i})$ and $h_{\lambda}(\mathbf{{G}}_{i})$:
\begin{equation}
\mbox{{logit}}(\pi_{i})=\mathbf{{X}}_{i}^{\top}\boldsymbol{{\beta}}_{\pi}+h_{\pi}(\mathbf{{G}}_{i})\label{eq:pi-linpred}
\end{equation}

\begin{equation}
\mbox{{log}}(\lambda_{i})=\mathbf{{X}}_{i}^{\top}\boldsymbol{{\beta}}_{\lambda}+h_{\lambda}(\mathbf{{G}}_{i})\label{eq:lda-linpred}
\end{equation}
Hence in our genetic testing paradigm we are interested in testing
$H_{0}:h_{\lambda}(\cdot)=h_{\pi}(\cdot)=0$. If $h_{\pi}(\cdot)\ne0,$
we say that genotypes are associated with the structural zeros, if
$h_{\lambda}(\cdot)\ne0$, we say that genotypes are associated with
the mean of the Poisson component of the mixture. For the purposes
of this paper, we assume that $h_{\pi}(\mathbf{{G}}{}_{i})=\mathbf{\Psi({G}}_{i})^{{\rm {\top}}}\boldsymbol{{\gamma}}_{\pi}$
and $h_{\lambda}(\mathbf{{G}}_{i})=\mathbf{\Psi({G}}_{i})^{{\rm {\top}}}\boldsymbol{{\gamma}}_{\lambda}$,
for a given set of basis functions $\mathbf{\Psi(\cdot})$. In practice,
$\mathbf{\Psi(\cdot})$ can be set to identity, leading to a testing
of linear effects or pre-specified using non-linear basis functions.
An alternative strategy to choose $\mathbf{\Psi(\cdot})$ is through
kernel principal component analysis as suggested in \citet{shen2016}.
More specifically, for a given kernel function and the observed data
on $\mathbf{{G}}{}_{i}$, one may estimate the basis functions associated
with the corresponding reproducible kernel Hilbert space by constructing
the kernel matrix for the observed sample and obtaining the leading
eigenvectors as $\mathbf{\Psi({G}}_{i})$. With a given $\mathbf{\Psi(\cdot})$,
testing the hypotheses of $H_{0}$ is then equivalent to testing $H_{0}:\boldsymbol{{\gamma}_{\pi}=}(\gamma_{\pi1},...,\gamma_{\pi K})^{{\rm {\top}}}=\boldsymbol{{\gamma}_{\lambda}=}(\gamma_{\lambda1},...,\gamma_{\lambda K})^{{\rm {\top}}}=0$.

\subsection{Test statistic}

To overcome the potential high dimensionality in $\mathbf{\Psi({G}}_{i}$)
and leverage the correlation among the SNPs, we propose to impose
a working assumption that ${\gamma_{\iota1},...,\gamma_{\iota K}}$
are independent and identically distributed random variables with
mean 0 and variance $\tau_{\iota}^{2}$, for $\iota\in\{\pi,\lambda\}$;
and derive a score test for the variance components $\tau_{\pi}$
and $\tau_{\lambda}$. Under this set-up, testing $H_{0}$ is now
translated into testing $H_{0}:\tau_{\pi}=\tau_{\lambda}=0$. The
variance component score test paradigm built on the random effects
working assumptions attains statistical efficiency gain essentially
by taking advantage of the correlation within $G$ to reduce the degrees
of freedom. These random effects working assumptions are merely a
convenience in developing the test statistic; they are not required
for the validity of the test. By using a score test we also have the
convenience of only needing to fit the null model where $\mbox{{logit}}(\pi_{0,i})=\mathbf{{X}}_{i}^{\top}\boldsymbol{{\beta}}_{0,\pi}$,
and $\mbox{{log}}(\lambda_{0,i})=\mathbf{{X}}_{i}^{\top}\boldsymbol{{\beta}}_{0,\lambda}$,
so that the model fit can be accomplished with standard maximum-likelihood
fixed-effects ZIP regression software. This means it is computationally
feasible to test a large number of SNP sets, even if, as with our
method, resampling is necessary to obtain the distribution of the
test statistic. 

To form the test statistic, we write $\gamma_{\iota k}=\tau_{\iota}\varepsilon_{\iota k}$
so that $E(\varepsilon_{\iota k})=0$, and $Var(\varepsilon_{\iota k})=1$,
and all covariances \textbf{$Cov(\varepsilon_{\iota k},\varepsilon_{\iota'k'})=0$}
when $\iota\ne\iota'$ or $k\ne k'$. Then for $\iota\in\left\{ \pi,\lambda\right\} $,
we define the score test statistic associated with $\tau_{\iota}$
as

\[
\hat{Q}_{\iota}=E_{\varepsilon}\left[\left.\left\{ \left.\frac{\partial\log\prod_{i=1}^{n}\mathcal{L}(\pi_{i},\lambda_{i}|\mathbf{D}_{i})}{\partial\tau_{\iota}}\right|_{\tau_{\pi}=\tau_{\lambda}=0}\right\} ^{2}\right|\mathbb{{D}}\right]=n\hat{\mathbf{{S}}}_{\iota}^{\top}\hat{\mathbf{{S}}}_{\iota}=\left\Vert \sqrt{n}\hat{\mathbf{{S}}}_{\iota}\right\Vert _{2}^{2},\quad\mbox{where\ }\hat{\mathbf{{S}}}_{\iota}=n^{-1}\sum_{i=1}^{n}\hat{r}_{\iota,i}\mathbf{\Psi({G}}_{i}).
\]
Here, $\hat{{r}}_{\pi,i}$ and $\hat{{r}}{}_{\lambda,i}$ can be thought
of as a covariate-adjusted scalar residual that is typical of a score
statistic. Explicitly, 

\begin{equation}
\hat{r}_{\pi,i}\equiv r_{\pi}(Y_{i},\mathbf{{X}}_{i}^{\top}\boldsymbol{\widehat{{\beta}}}_{0,\pi},\mathbf{{X}}_{i}^{\top}\boldsymbol{\widehat{{\beta}}}_{0,\lambda})\equiv I(Y_{i}=0)\frac{\hat{\pi}_{0,i}(1-\hat{\pi}_{0,i})(1-exp(-\hat{\lambda}_{0,i}))}{1-\hat{\pi}_{0,i}(1-exp(-\hat{\lambda}_{0,i}))}-I(Y_{i}>0)(1-\hat{\pi}_{0,i})\label{eq:PiResid}
\end{equation}

and 
\begin{equation}
\hat{r}_{\lambda,i}\equiv r_{\lambda}(Y_{i},\mathbf{{X}}_{i}^{\top}\boldsymbol{\widehat{{\beta}}}_{0,\pi},\mathbf{{X}}_{i}^{\top}\boldsymbol{\widehat{{\beta}}}_{0,\lambda})\equiv I(Y_{i}=0)\frac{\hat{\pi}_{0,i}\hat{\lambda}_{0,i}exp(-\hat{\lambda}_{0,i})}{1-\hat{\pi}_{0,i}(1-exp(-\hat{\lambda}_{0,i}))}-I(Y_{i}>0)(Y_{i}-\hat{\lambda}_{0,i}),\label{eq:LdaResid}
\end{equation}
where $\mbox{{logit}}\hat{\pi}_{0,i}=\mathbf{{X}}_{i}^{\top}\boldsymbol{\widehat{{\beta}}}_{0,\pi}$,
$\log\hat{\lambda}_{0,i}=\mathbf{{X}}_{i}^{\top}\boldsymbol{\widehat{{\beta}}}_{0,\lambda}$,
and $\widehat{{\boldsymbol{\beta}}_{0}}=(\boldsymbol{\widehat{{\beta}}}_{0,\pi}^{\top},\boldsymbol{\widehat{{\beta}}}_{0,\lambda}^{\top})^{\top}$
is the maximum likelihood estimator of the covariate effects $\boldsymbol{{\beta}}_{0}=(\boldsymbol{{\beta}}_{0,\pi}^{\top},\boldsymbol{{\beta}}_{0,\lambda}^{\top})^{\top}$under
$H_{0}$. Note that $\widehat{E(Y_{i})}=\hat{\pi}_{0,i}\hat{\lambda}_{i}$,
$\widehat{P(Y_{i}>0)}=\hat{\pi}_{0,i}(1-exp(-\hat{\lambda}_{0,i}))$,
so that $\hat{r}_{\pi,i}=I(Y_{i}=0)(1-\hat{\pi}_{0,i})\frac{\widehat{P(Y_{i}>0)}}{\widehat{P(Y_{i}=0)}}-I(Y_{i}>0)(1-\hat{\pi}_{0,i})$
and $\hat{r}_{\lambda,i}=I(Y_{i}=0)\frac{\widehat{E(Y_{i})}exp(-\hat{\lambda}_{0,i})}{\widehat{P(Y_{i}=0)}}-I(Y_{i}>0)(Y_{i}-\hat{\lambda}_{0,i})$.
For $\iota\in\left\{ \pi,\lambda\right\} $, the statistic $\hat{Q}_{\iota}$
can be interpreted as the $L_{2}$ norm of an empirical covariance
between the residuals $\left\{ \hat{r}_{\iota,i}\right\} $ estimated
under the null of no genetic effect and the transformed genotypes
$\left\{ \mathbf{\Psi}(\mathbf{G}_{i})\right\} $. 

In Appendix 1, we show that under $H_{0}$, the covariances $\hat{\mathbf{{S}}}_{\pi}$
and $\hat{\mathbf{{S}}}_{\lambda}$ converges to zero in probability
and $\sqrt{{n}}\hat{\mathbf{{S}}}\equiv\sqrt{{n}}(\hat{\mathbf{{S}}}_{\pi}^{\top},\hat{\mathbf{{S}}}_{\lambda}^{\top})^{\top}$
converges in distribution to a multivariate normal with covariance
matrix $\boldsymbol{{\Sigma}}=\left[\begin{array}{cc}
\boldsymbol{{\Sigma}}_{\pi,\pi} & \boldsymbol{{\Sigma}}_{\pi,\lambda}\\
\boldsymbol{{\Sigma}}_{\pi,\lambda}^{\top} & \boldsymbol{{\Sigma}}_{\lambda,\lambda}
\end{array}\right]$, where $\boldsymbol{{\Sigma}}_{\iota,\iota'}=\mbox{cov}(\boldsymbol{\Phi}_{\iota,i},\boldsymbol{\Phi}_{\iota',i})$,
$\boldsymbol{\Phi}_{\iota,i}={r_{\iota,i}}\mathbf{\Psi({G}}_{i})+\rho_{\iota,\pi}\mathbb{I}_{\pi}^{-1}\mathbf{U}_{\pi,i}+\rho_{\iota,\lambda}\mathbb{I}_{\lambda}^{-1}\mathbf{U}_{\lambda,i}$,
$\mathbf{U}_{\iota,i}$ and $\mathbb{I}_{\iota}$ are the respective
score and information matrix for $\beta_{\iota}$, $\rho_{\iota,\pi}=E\left[\dot{{r}}_{2\iota}(Y_{i},\mathbf{{X}}_{i}^{\top}\boldsymbol{{\beta}}_{0,\pi},\mathbf{{X}}_{i}^{\top}\boldsymbol{{\beta}}_{0,\lambda})\mathbf{\Psi({G}}_{i})\mathbf{{X}}_{i}^{\top}\right]$,
$\rho_{\iota,\lambda}=E\left[\dot{{r}}_{3\iota}(Y_{i},\mathbf{{X}}_{i}^{\top}\boldsymbol{{\beta}}_{0,\pi},\mathbf{{X}}_{i}^{\top}\boldsymbol{{\beta}}_{0,\lambda})\mathbf{\Psi({G}}_{i})\mathbf{{X}}_{i}^{\top}\right]$,
and $\dot{{r}}_{k\iota}=\partial r_{\iota}(x_{1},x_{2},x_{3})/\partial x_{k}$.
It follows (see Appendix 1) that $\hat{Q}_{\iota}$ converges in distribution
to a mixture of $\chi_{1}^{2}$ distributions with mixing parameters
being the eigenvalues of $\boldsymbol{{\Sigma}}_{\iota,\iota}$, for
$\iota\in{\pi,\lambda}$.

To form an omnibus test for $H_{0}$ combining information from the
two score statistics $\hat{Q}_{\pi}$ and $\hat{Q}_{\lambda}$, several
general approaches for combing p-values can be employed, for example
a min-p approach \citep{huang2014,won2009choosing}, or a Fisher's
method approach \citet{fisher1925statistical}. In this case we calculate
the separate p-values $\hat{p}_{\pi}$, and $\hat{p}_{\lambda}$ for
$\pi$ and for $\lambda,$ and obtain an estimate of their joint distribution
via perturbation resampling so that a size $\alpha$ test can be computed
for the statistics $\hat{p}_{m}=min(\hat{p}_{\pi},\hat{p}_{\lambda})$,
and $\hat{p}_{F}=log(\hat{p}_{\pi})+log(\hat{p}_{\lambda})$, as described
below. These approaches have the advantage of putting the two statistics
$\hat{Q}_{\pi}$ and $\hat{Q}_{\lambda}$ on the same scale via p-value
for combination. Alternatively, we may rescale $\hat{\mathbf{{S}}}_{\pi}$
and $\hat{\mathbf{{S}}}_{\lambda}$ as $\hat{\sigma}{}_{\pi}^{-1}\hat{\mathbf{{S}}}_{\pi}$
and $\hat{\sigma}_{\lambda}^{-1}\hat{\mathbf{{S}}}_{\lambda}$ and
then construct $\hat{Q}_{\pi,\lambda}=n\sum_{\iota\in\left\{ \pi,\lambda\right\} }\hat{\sigma}_{\iota}^{-2}\hat{\mathbf{{S}}}_{\iota}^{\top}\hat{\mathbf{{S}}}_{\iota}=\sum_{\iota\in\left\{ \pi,\lambda\right\} }\hat{\sigma}_{\iota}^{-2}\hat{{Q}_{\iota}}$,
where $\hat{\sigma}_{\iota}^{2}=\mbox{{trace}}(\widehat{\boldsymbol{\Sigma}}_{\iota,\iota})$
and $\widehat{\boldsymbol{\Sigma}}_{\iota,\iota}$ is the estimated
$\boldsymbol{{\Sigma}}_{\iota,\iota}$.

\subsection{Resampling Methods}

To approximate the null distribution of the test statistics in practice,
we employ a resampling method to approximate $\boldsymbol{{\Sigma}}$
as well as the joint distribution of $\hat{p}_{\pi}$and $\hat{p}_{\lambda}$under
$H_{0}$. Specifically we generate $n$ independent samples $\mathbf{{V}}=\{V_{i}\}_{i=1}^{n}$
sampled from some fixed distribution with $E[V_{i}]=1$, $Var[V_{i}]=1$,
such as $V_{i}\sim Exponential(1)$. For a given set of weights$\mathbf{{V}}^{(b)}=\{V_{i}^{(b)}\}_{i=1}^{n}$,
we obtain the perturbed score vectors $\hat{\mathbf{{S}}}^{(b)}=(\hat{\mathbf{{S}}}_{\pi}^{(b)\top},\hat{\mathbf{{S}}}_{\lambda}^{(b)\top})^{\top}$,
where 

\begin{equation}
\hat{\mathbf{{S}}}_{\iota}^{(b)}=(\sum_{i=1}^{n}V_{i}^{(b)})^{-1}\sum_{i=1}^{n}\hat{r}_{\iota,i}^{(b)}\mathbf{\Psi({G}}_{i})V_{i}^{(b)},\quad\hat{r}_{\iota,i}^{(b)}\equiv r_{\iota}(Y_{i},\mathbf{{X}}_{i}^{\top}\boldsymbol{\widehat{{\beta}}}_{0,\pi}^{(b)},\mathbf{{X}}_{i}^{\top}\boldsymbol{\widehat{{\beta}}}_{0,\lambda}^{(b)}),\quad\label{eq:Resid-p}
\end{equation}
and $\boldsymbol{\widehat{{\beta}}}_{0}^{(b)}=\left(\boldsymbol{\widehat{{\beta}}}_{0,\pi}^{(b)},\boldsymbol{\widehat{{\beta}}}_{0,\lambda}^{(b)}\right)$
is obtained by fitting a weighted ZIP regression under the null using
weights $\mathbf{{V}}^{(b)}$. It can be shown that, under suitable
regularity conditions, $\sqrt{n}\left(\mathbf{{\hat{S}}}^{(b)}-\mathbf{{\hat{S}}}\right)\left|\mathbf{D}\right.\overset{\mathcal{D}}{\rightarrow}MVN(\boldsymbol{0},\boldsymbol{\Sigma})$
\citep{parzen1994,kline2012}. Hence, for some large number $B$ of
repetitions of the resampling procedure, we use the empirical distribution
$\left\{ \hat{Q}_{\iota}^{(b)}\equiv n(\hat{\mathbf{{S}}}_{\iota}^{(b)}-\hat{\mathbf{{S}}}_{\iota})^{\top}(\hat{\mathbf{{S}}}_{\iota}^{(b)}-\hat{\mathbf{{S}}}_{\iota})\right\} _{b=1}^{B}$
to approximate the asymptotic distribution of $\hat{Q}_{\iota}$.
In order to calculate $p$-values with greater precision than $1/B$
we use Imhof's method \citep{imhof1961} to obtain the distribution
of the quadratic form $\hat{Q}_{\iota}$, under the established multivariate
normality on the half-statistic $\hat{\mathbf{{S}}}_{\iota}$, via
the eigendecomposition of the empirical covariance of the resampling
distribution $\left\{ (\hat{\mathbf{{S}}}_{\iota}^{(b)}-\hat{\mathbf{{S}}}_{\iota})\right\} _{b=1}^{B}$
. This procedure is facilitated by the $R$ package `CompQuadForm'
\citet{duchesne2010quadratic}.

To calculate a $p$-value for the omnibus test, appropriately controlling
type I error in the presence of correlation between $\hat{p}_{\pi}$
and $\hat{p}_{\lambda}$, we take advantage of the same resampling
procedure described above. For each $\hat{\mathbf{{S}}}_{\pi}^{(b)}$
and $\hat{\mathbf{{S}}}_{\lambda}^{(b)}$, $b=1,...,B$, we use Imhof's
method to calculate a corresponding $\hat{p}_{\pi}^{(b)}$ and $\hat{p}_{\lambda}^{(b)}$
. Define $\hat{p}_{m}=min(\hat{p}_{\pi},\hat{p}_{\lambda})$. Then
the null distribution of $\hat{p}_{m}$ can be approximated using
the resampling distribution$\{\hat{p}_{m}^{(b)}\}_{b=1}^{B}$ where
$\hat{p}_{m}^{(b)}=min(\hat{p}_{\pi}^{(b)},\hat{p}_{\lambda}^{(b)})$.
We can obtain a min-p $p$-value as $\hat{p}_{min}=\frac{1}{B}\sum_{b}I\left(\hat{p}_{m}^{(b)}\le\hat{p}_{m}\right)$.
The null distribution of $\hat{p}_{F}$ is similarly obtained, leading
to an alternative omnibus $p$-value: $\hat{p}_{Fisher}=\frac{1}{B}\sum_{b}I\left(\hat{p}_{F}^{(b)}\le\hat{p}_{F}\right)$.
In our simulations we show that both $\hat{p}_{min}$ and $\hat{p}_{Fisher}$
have appropriate size under the null. Heuristically, under the alternative
we expect $\hat{p}_{Fisher}$ to perform worse than $\hat{p}_{min}$
with respect to power in cases when the signal is either through $\pi$
or $\lambda$ but not both, and to outperform $p_{min}$ when an association
signal appears in both $\pi$ and $\lambda$. Indeed our simulations
confirm $\hat{p}_{Fisher}$ modestly outperforms $\hat{p}_{min}$
when the association signal is through both parameters.

\section{Results}

\subsection{Simulations}

\subsubsection{Simulation Setting}

We generated data from a zero-inflated Poisson model of the form \ref{eq:model}
with $\mbox{{logit}}(\pi_{i})=\alpha_{\pi}+\mathbf{X}_{i}^{\top}\boldsymbol{{\beta}}_{\pi}+\mathbf{G}_{Yi}^{\top}\boldsymbol{{\gamma}}_{\pi}$,
and $\log(\lambda_{i})=\alpha_{\lambda}+\mathbf{X}_{i}^{\top}\boldsymbol{{\beta}}_{\lambda}+\mathbf{G}_{Yi}^{\top}\boldsymbol{{\gamma}}_{\lambda}$.
We varied both the genetic data distribution and parameter settings
across five overall simulation scenarios, the first four of which
examine size of the test under the null. In setting 1), we fix the
sample size at $n=1000$ but considered simple cases where $\mathbf{{X}}$
is either absent or is independent of $\mathbf{{G}}$. In setting
2), we let $\mathbf{{X}}$ to be moderately correlated with $\mathbf{{G}}$
and let the sample size $n$ = 1000, 2000 and 4000. In setting 3),
we varied the correlation strength between $\mathbf{{X}}$ and $\mathbf{{G}}$,
up to a maximum pairwise correlation of approximately $0.9$ to investigate
the impact of correlation on the results. In setting 4), we simulated
data with overdispersion by including an unobserved covariate associated
with the outcome. This was to determine how robust our method is to
this important deviation from the assumptions of the specified model.
In setting 5), we generated data under the alternative, where the
outcome was associated with (i) both $\pi_{i}$ and $\lambda_{i}$
, (ii) only with $\pi_{i}$ , and (iii) only with $\lambda_{i}$ ,
in order investigate how power depends on the source of the signal.
Here, the covariates $\mathbf{{X}}$ were generated independent of
$\mathbf{{G}}$. 

Across all scenarios to obtain genetic marker data $\mathbf{G}_{i}$
we used HAPGEN2 \citep{su2011} to generate haplotypes from 1000 genomes
phase one data, CEU panel, using two representative SNP sets. First,
from the APOE gene, known to be associated with our neuritic plaque
burden in our example, we used the 8 SNPs in the set \{rs10119, \textbf{rs429358},
\textbf{rs7412}, 19-50106239, rs445925, rs483082, rs59325138, \textbf{rs438811}\},
and second from the CD33 gene, we used the 22 SNPs in the set \{rs273637,
rs273638, rs273639, \textbf{rs273640}, rs1399837, rs3826656, rs1710398,
rs1697553, 19-56419774, rs2459141, 19-56420453, rs7245846, \textbf{rs34813869},
rs1354106, rs35112940, \textbf{rs10409348}, rs273652, rs1697531, rs169275,
rs273649, rs273648, rs273646\}. In either case under the alternative,
3 SNPs $G_{Yi}$ (in bold) were associated with the outcome under
the true data generation model. The APOE gene has a low-moderate LD
across all selected SNPs while the CD33 gene has two independent haploblocks
with strong LD. We used these specific SNP sets to account for the
effect of differing linkage disequilibrium (LD) structures (correlation)
and genotype dimension on the performance of our method when applied
to various genes. 

In each scenario we compared our method to the multivariate Wald test
from three models: the zero-inflated poisson, the zero-inflated negative
binomial, and a simple Poisson model using the robust Huber-White
(HW) variance. For each competing model except the single-parameter
Poisson model, we performed multivariate Wald tests on $\widehat{\boldsymbol{\gamma}}_{\pi}$
and $\widehat{\boldsymbol{\gamma}}_{\lambda}$ separately, as well
as the combined test of the vector concatenation $(\widehat{\boldsymbol{\gamma}}_{\pi}^{\top},\widehat{\boldsymbol{\gamma}}_{\lambda}^{\top})^{\top}$.
For each Wald test, because Poisson, ZIP, and ZINB models failed to
achieve convergence with maximum likelihood due to the collinearity
in these SNP sets, in each round of our simulation we performed an
LD pruning on the matrix $\mathbf{G}$, with SNPs chosen so that no
pairwise correlation was above 0.99. For APOE, this LD pruning resulted
in no removal of SNPs, however for CD33, this reduced the number of
SNPs in the model from 22 to approximately 12, and potentially varied
depending on the data.

To specify covariates, for those scenarios where $\mathbf{X}_{i}$
is independent of $\mathbf{G}_{i}$, we set $\mathbf{X}_{Ind,i}=(X_{1i},X_{2i},X_{3i},X_{4i},X_{5i})^{\top}$,
where each $X_{ki}$ is a complex linear combination of independent
binomial and normal random variables, with pairwise correlations that
ranged from approximately 0.1 to 0.9. The complete specification is
shown in the Appendix. When $\mathbf{X}_{i}$ is dependent on $\mathbf{G}_{i}$,
we set $\mathbf{X}_{i,Dep}=\rho_{XG}\mathbf{A}\mathbf{G}_{Yi}+\mathbf{X}_{Ind,i}$,
where the $\mathbf{X}:\mathbf{G}$ association is through a matrix
$\mathbf{A}$ shown in the Appendix, and $\rho_{XG}$ is a scalar
correlation strength parameter chosen to be 0.25, 0.5, or 1, representing
a mild, moderate, or strong correlation, respectively. . When $\rho_{XG}=$0.25,
0.5 and 1, the maximal pairwise correlations in the $\mathbf{X}:\mathbf{G}$
correlation matrix were approximately 0.23, 0.44 and 0.64 for APOE
and 0.58, 0.84, and 0.92 for CD33 respectively. 

For all settings, $\alpha_{\pi}=1.5$, $\alpha_{\lambda}=1.3$, $\boldsymbol{\beta}_{\pi}=(0.75,0.5,0.25,1.0,1.0)$,
$\boldsymbol{\beta}_{\lambda}=(0.25,0.5,0.75,1.0,1.0)$. For settings
where $\mathbf{G}_{Yi}$ is associated with the outcome, for APOE,
we set $\boldsymbol{\gamma}_{\pi}=(0.066,0.33,0.66)$, and $\boldsymbol{\gamma}_{\lambda}=(0.11,0.055,0.011)$,
while for CD33, we set $\boldsymbol{\gamma}_{\pi}=(0.036,0.18,0.36)$
, and $\boldsymbol{\gamma}_{\lambda}=(0.06,0.03,0.006)$. Depending
on the gene, the dimension of $\mathbf{G}_{i}$ was either $p=8$
or $p=22$, but in each case there were 3 SNPs $\mathbf{G}_{Yi}$
associated with $Y_{i}$ under the alternative, while the rest were
not included in the true association model. 

For setting 4) in which the model is mis-specified due to overdispersion,
the true data generating mechanism is parameterized by $\mbox{{logit}}(\pi_{i})=\alpha_{\pi}+\mathbf{X}_{i}^{\top}\boldsymbol{{\beta}}_{\pi}+\mathbf{G}_{Yi}^{\top}\boldsymbol{{\gamma}}_{\pi}$,
(as above), but $\log(\lambda_{i})=\alpha_{\lambda}+\mathbf{X}_{i}^{\top}\boldsymbol{{\beta}}_{\lambda}+\mathbf{G}_{Yi}^{\top}\boldsymbol{{\gamma}}_{\lambda}+\varepsilon_{\lambda,i}$
where $\varepsilon_{\lambda,i}\sim Normal(0,0.3^{2})$. 

All testing was performed using the linear kernel test statistic with
$\boldsymbol{{\Psi}}(\mathbf{{G}})=\mathbf{{G}}$. In each scenario,
the nominal size of the type I error is $\alpha=.05$. Unless otherwise
stated, the simulated data had sample size $n=1000$.

\subsubsection{Simulation Results}

We first compare our proposed variance component (VC) tests of the
individual parameters $\pi$ and for $\lambda$ described above (here
denoted $\mbox{{VC}}{}_{\pi}$ and $\mbox{{VC}}{}_{\lambda}$) with
the corresponding Wald tests from the zero-inflated Poisson model
and the zero-inflated negative binomial model (denoted $\mbox{{Wald}}_{\pi,ZIP}$,
$\mbox{{Wald}}_{\pi,ZINB}$, $\mbox{Wald}{}_{\lambda,ZIP}$ and $\mbox{{Wald}}_{\lambda,ZINB}$).
Additionally we compared six tests of the global null: three methods
of combining our tests on $\pi$ and $\lambda$, namely the min-p
approach ($\mbox{{VC}}_{min}$), the Fisher's method approach ($\mbox{{VC}}_{Fisher}$),
and the variance-standardized approach ($\mbox{{VC}}_{std}$) based
on the statistic $\hat{Q}_{\pi,\lambda}$ proposed in Section 2, along
with the Wald test using the Huber-White robust variance from the
Poisson model ($\mbox{{HW}}_{Pois}$) and the Wald tests of combined
$\pi$ and $\lambda$ vector in the ZIP and ZINB settings ($\mbox{{Wald}}_{ZIP}$,
$\mbox{{Wald}}_{ZINB}$).

\subsubsection{Size of the tests under correct model specification }

\begin{figure}
\includegraphics[bb=50bp 60bp 270bp 240bp,clip,width=6cm]{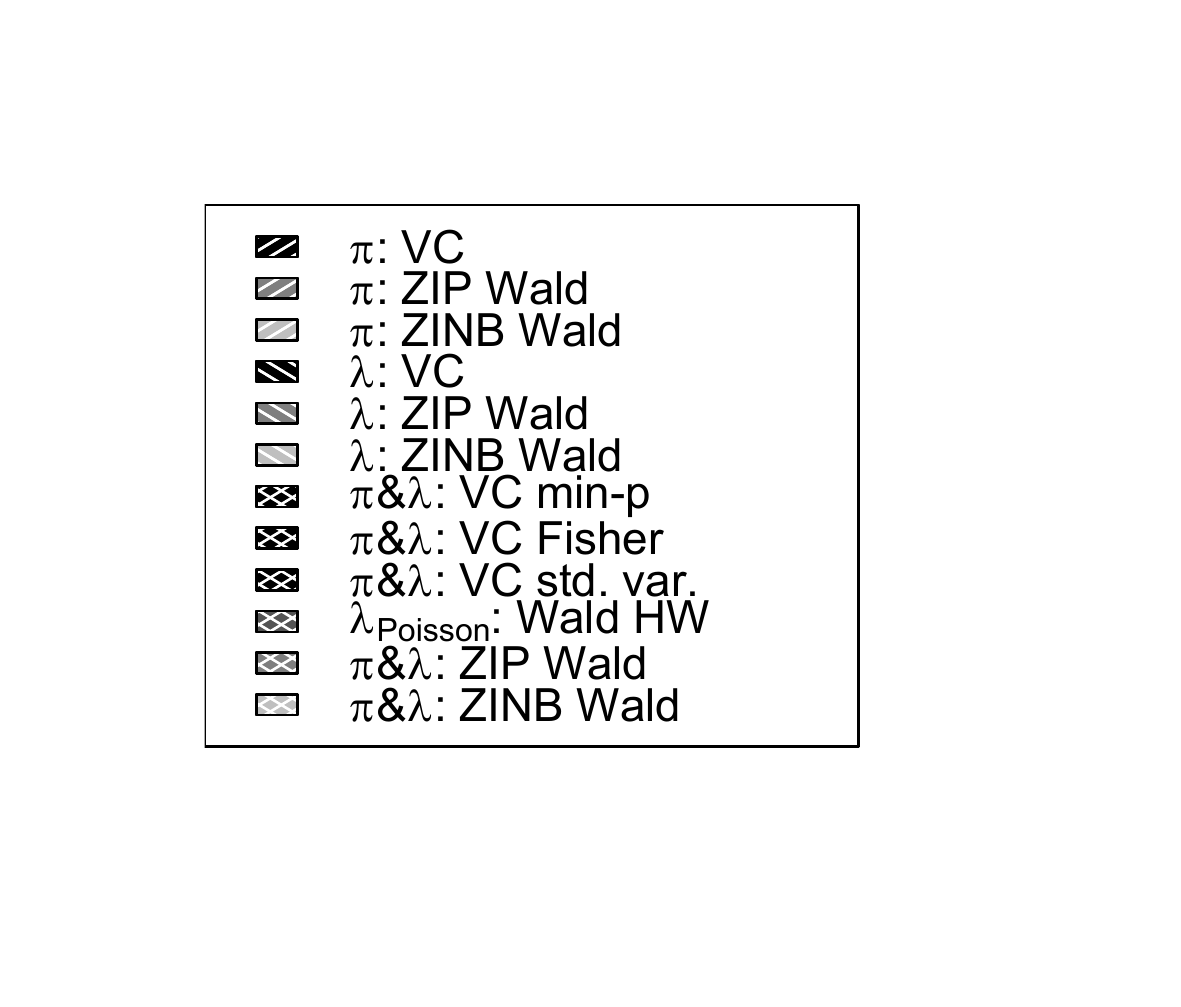}

\caption{Legend for following plots of simulation results. \label{tab:PlotLegend}}
\end{figure}

\begin{figure}
\includegraphics[width=8cm]{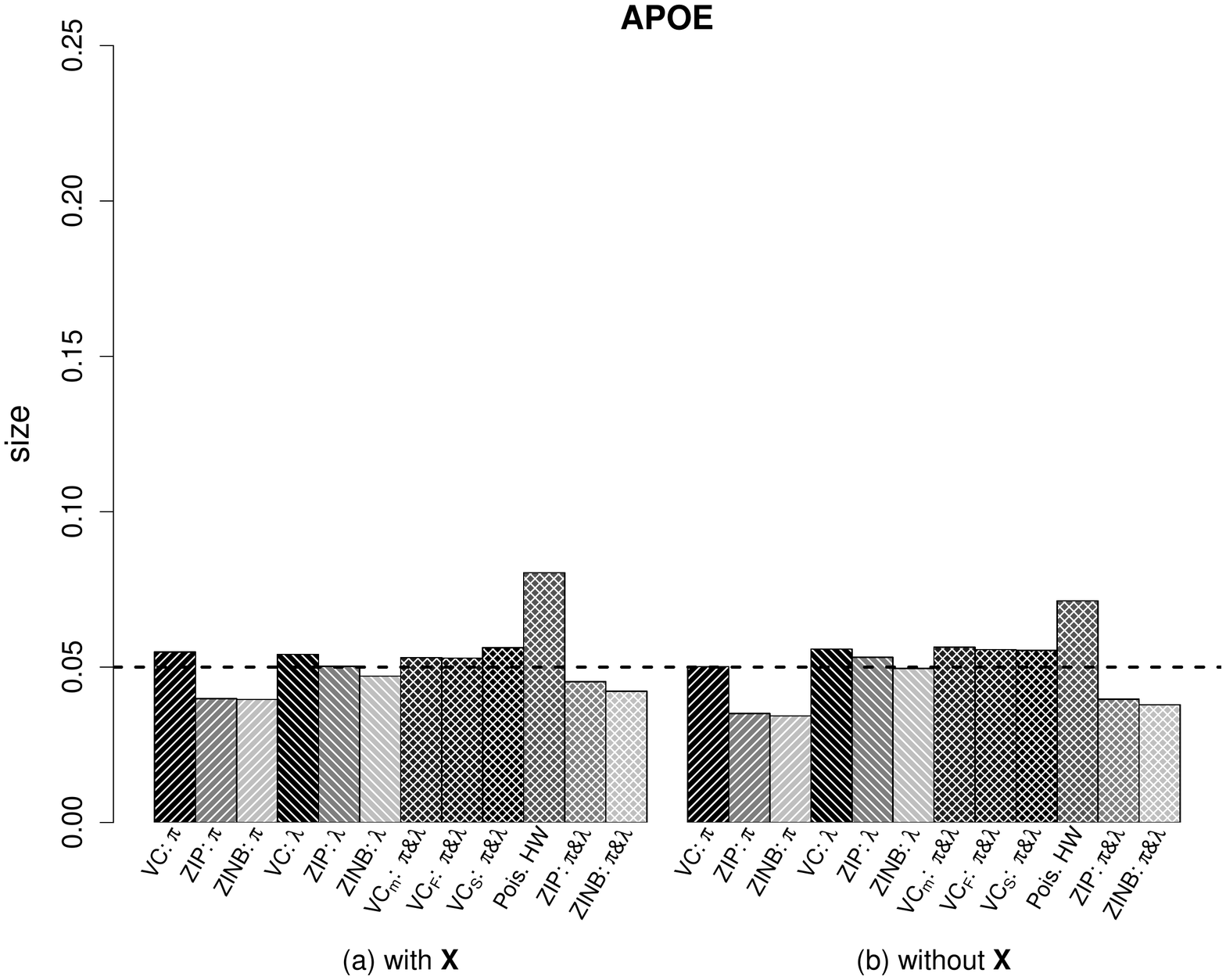}\includegraphics[width=8cm]{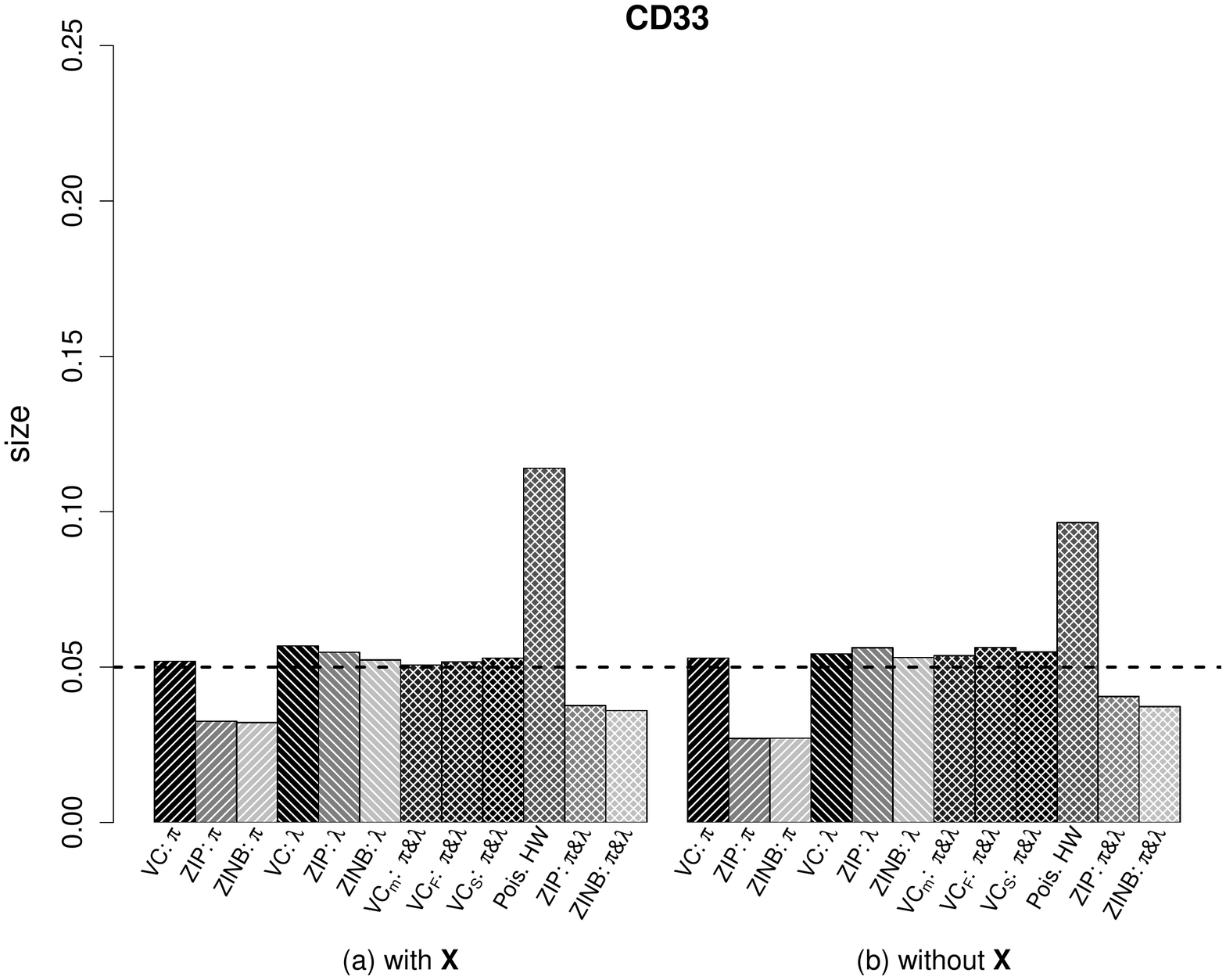}

\caption{Empirical Sizes of different testing procedures for Setting 1) (a)
without$\mathbf{{X}}$; and (b) with$\mathbf{{X}}$.\label{fig:Size}}
\end{figure}

\begin{figure}
\includegraphics[width=8cm]{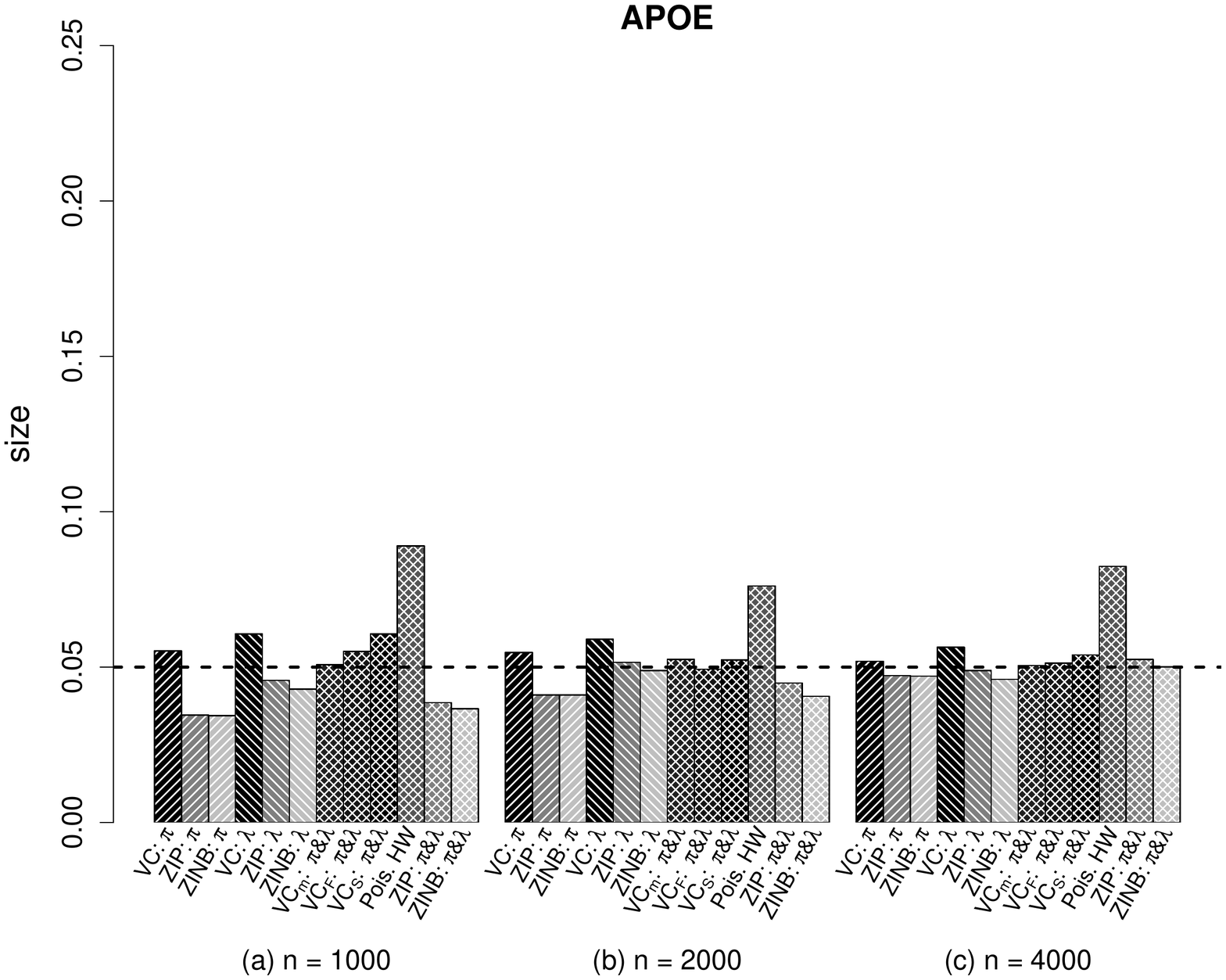}\includegraphics[width=8cm]{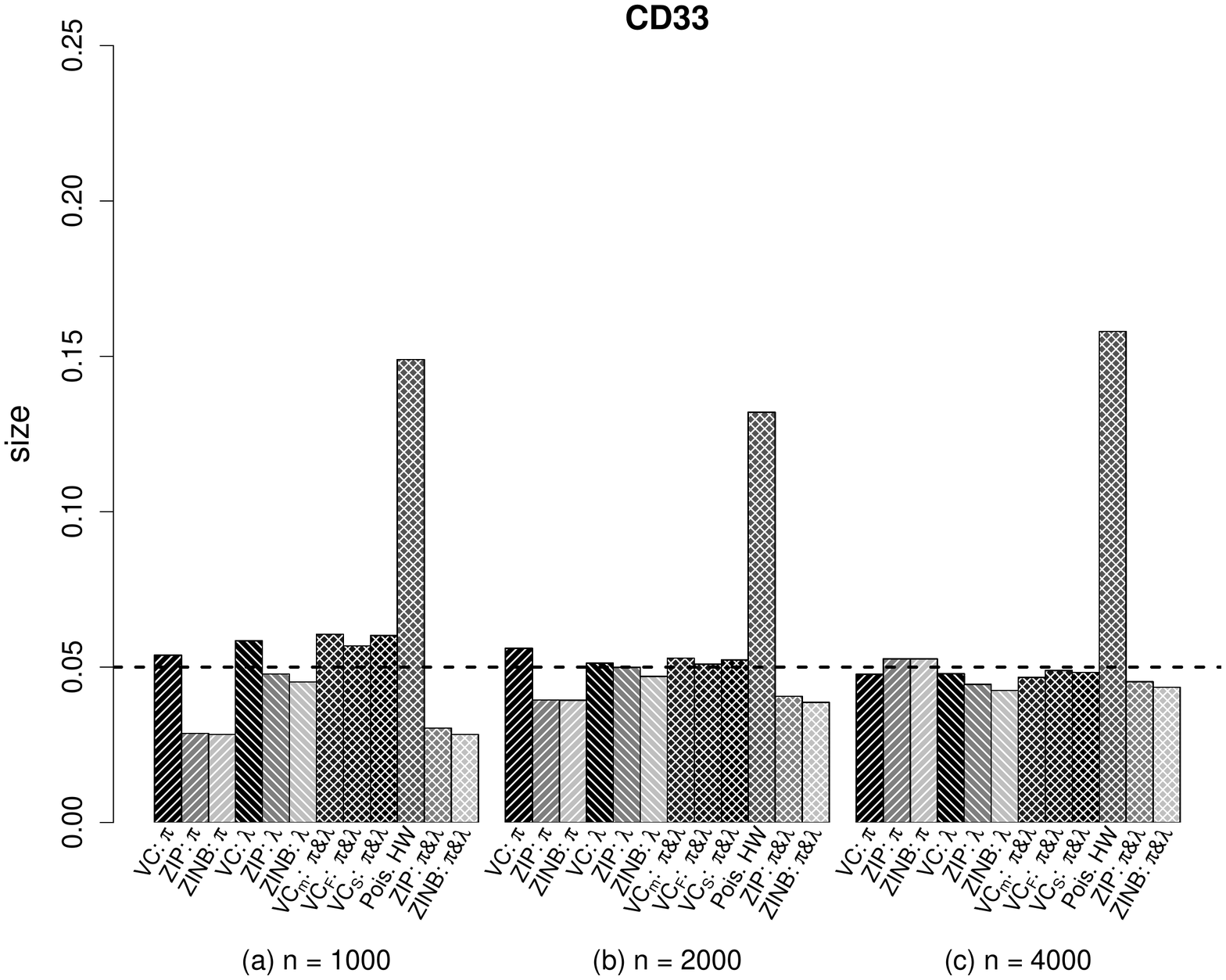}

\caption{Empirical sizes of different testing procedures for Setting 2) when
$n=$1000, 2000, and 4000.\label{fig:Size-N}}
\end{figure}
\begin{figure}
\includegraphics[width=8cm]{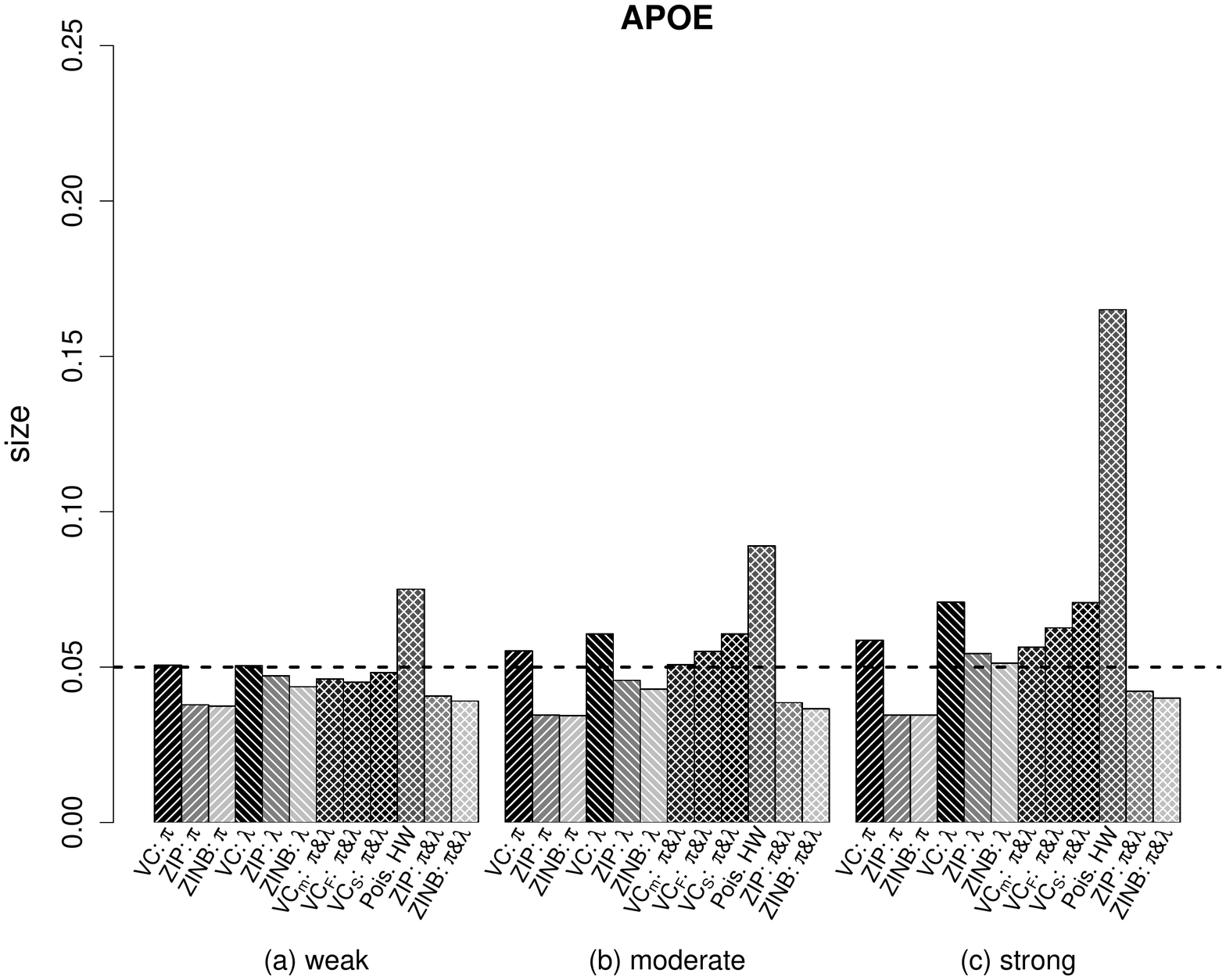}\includegraphics[width=8cm]{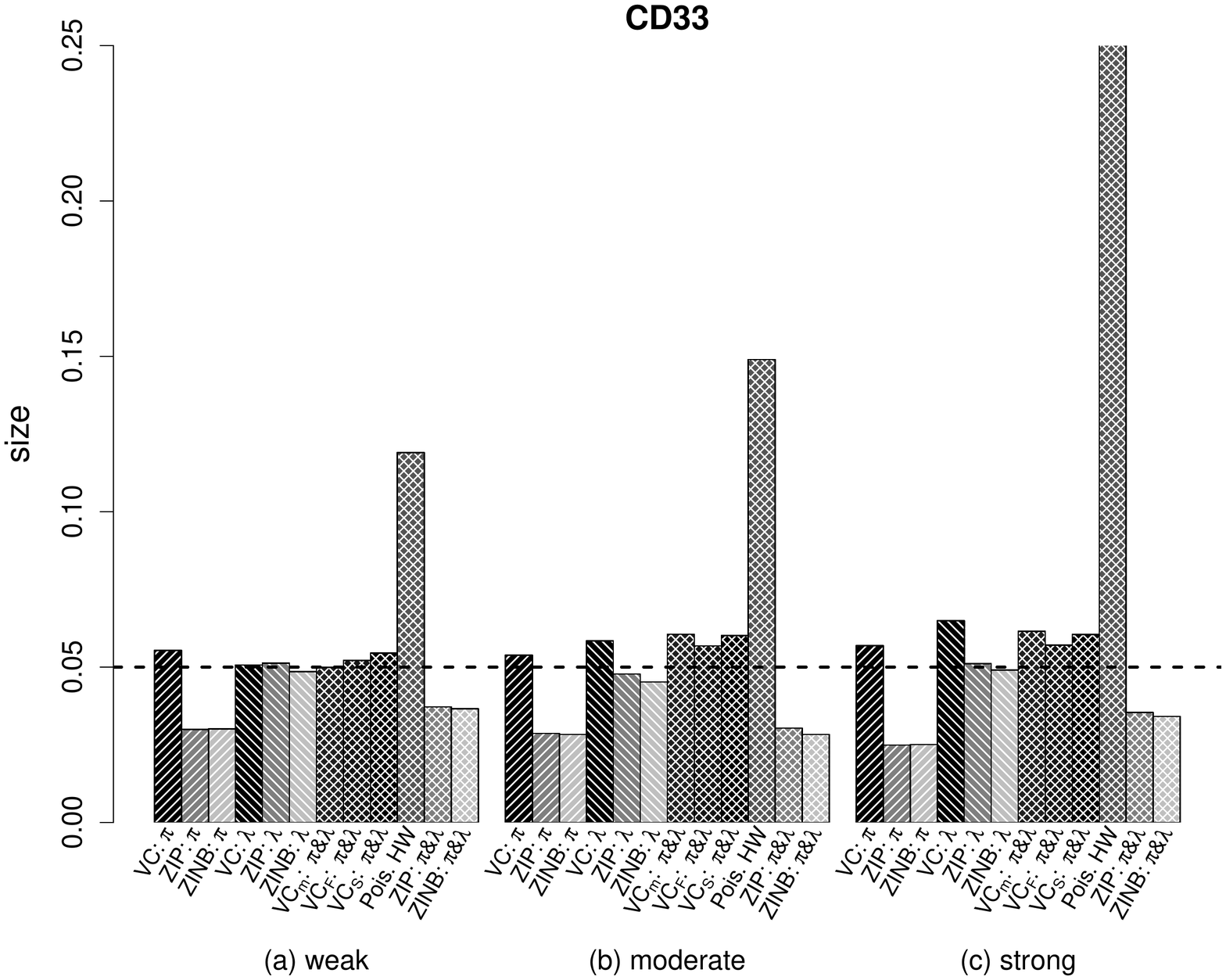}

\caption{Empirical sizes of different testing procedures for Setting 3) when
the correlation between $\mathbf{{X}}$ and $\mathbf{{G}}$ is weak,
moderate or strong. \label{fig:Size-Corr}}
\end{figure}
Settings 1) - 3) allow us to whether the size of the test is maintained
in finite sample under correct specification of the model but with
varying configurations for sample sizes and correlation structures.
Results summarized in Figures \ref{fig:Size}, \ref{fig:Size-N} and
\ref{fig:Size-Corr} suggest that type I errors of all methods except
for the HW procedure are well-controlled, with empirical sizes close
to the nominal $\alpha$ of 0.05 across all scenarios, under both
the APOE and CD33 LD structures. The HW procedure appears to yield
inflated type I errors for a majority of the settings and is thus
removed from subsequent power comparisons. It is worth noting that
the correlation between the covariates $\mathbf{X}_{i}$ and the tested
genetic markers $\mathbf{G}_{i}$ plays relatively little role in
affecting type I error of our method unless levels of correlation
are quite high, e.g. where the maximal element of the $\mathbf{X}:\mathbf{G}$
correlation matrix is above 0.8.

\subsubsection{Effect of mis-specification (overdispersion) on size of tests}

\begin{figure}
\includegraphics[width=8cm]{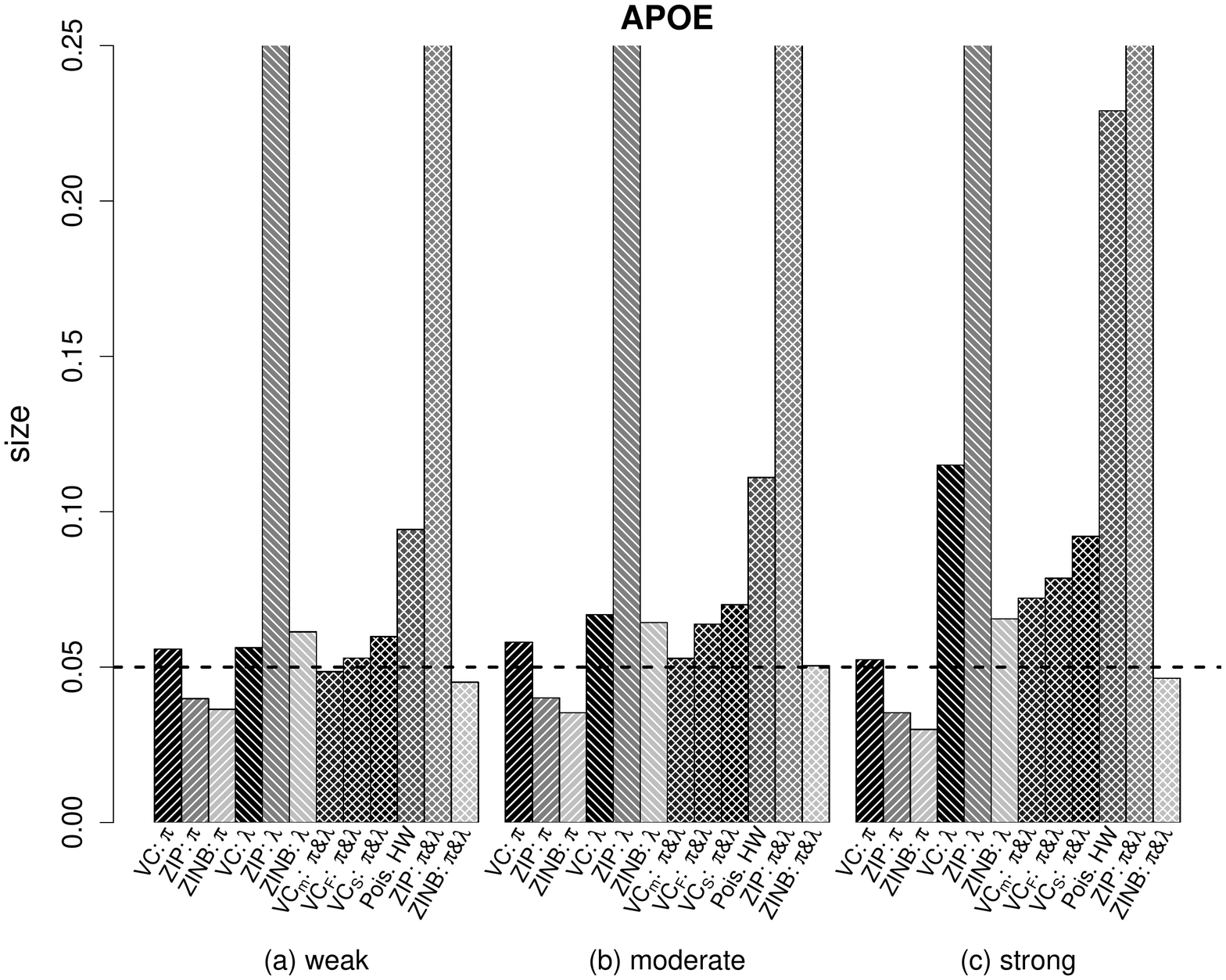}\includegraphics[width=8cm]{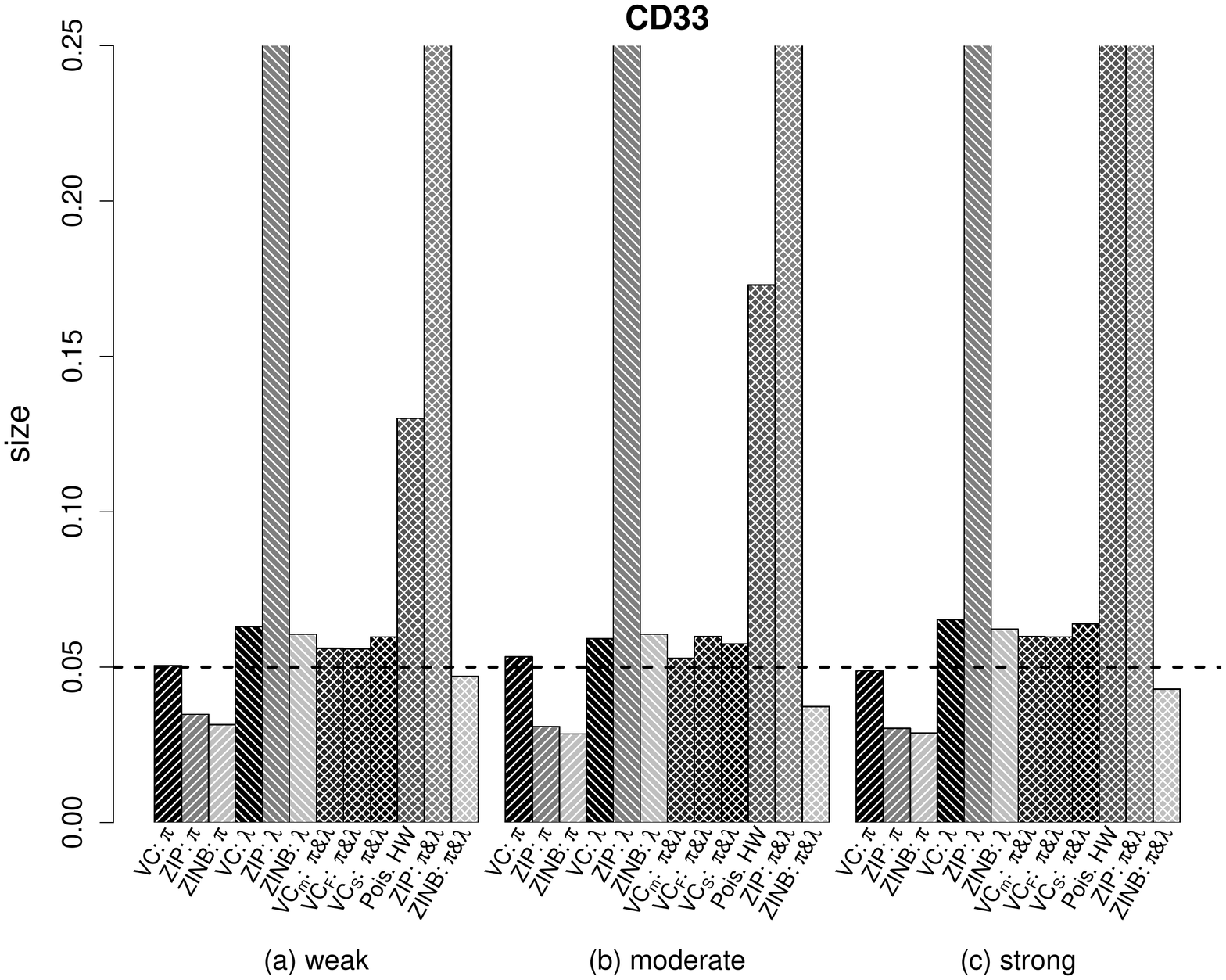}

\caption{Empirical sizes of different testing procedures for Setting 4) with
overdispersion and weak, moderate or strong $\mathbf{{X:G}}$ correlation.
\label{fig:Size-OD}}
\end{figure}

In this scenario we investigate the robustness of different testing
procedures in preserving type I errors in the presence of overdispersion
and hence model mis-specification. Specifically, as described above,
we included a Normally distributed covariate in the ZIP data generating
model for $\lambda$ that was not controlled for in the models we
used to derive the tests. We see in Figure \ref{fig:Size-OD} that
the type I error of our VC test was relatively robust to overdispersion,
but performance was compromised in APOE with increased $\mathbf{X}:\mathbf{G}$
correlation. As a comparison, the ZIP Wald test was badly compromised
in terms of type I error with the amount of overdispersion in this
setting, resulting in a sizes ranging from $0.4$ to $0.9$. Both
the Wald test based on ZINB and our VC tests only have slight inflation
of type I error in the presence of mis-specification due to overdispersion. 

\subsubsection{Power }

\begin{figure}
\includegraphics[width=8cm]{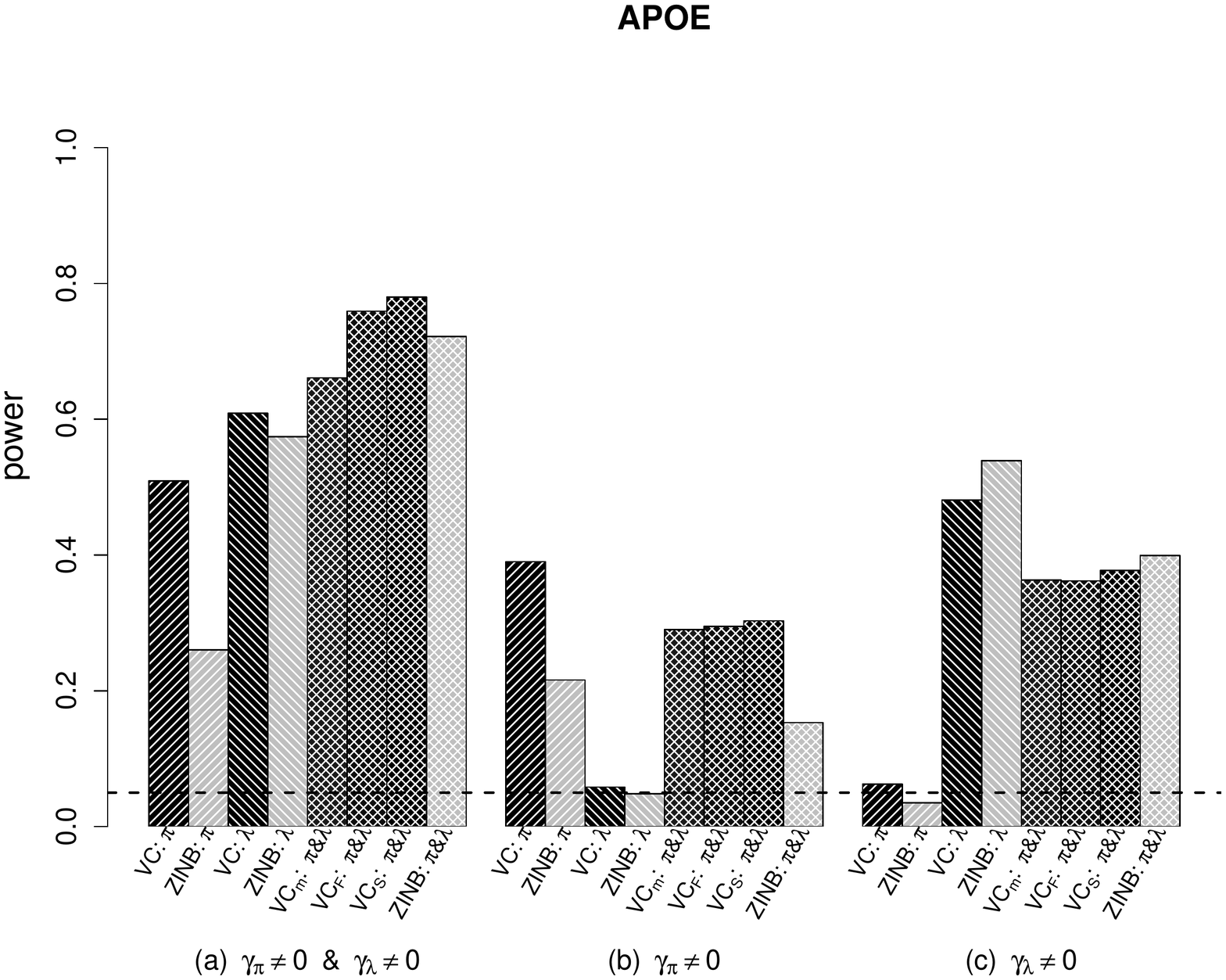}\includegraphics[width=8cm]{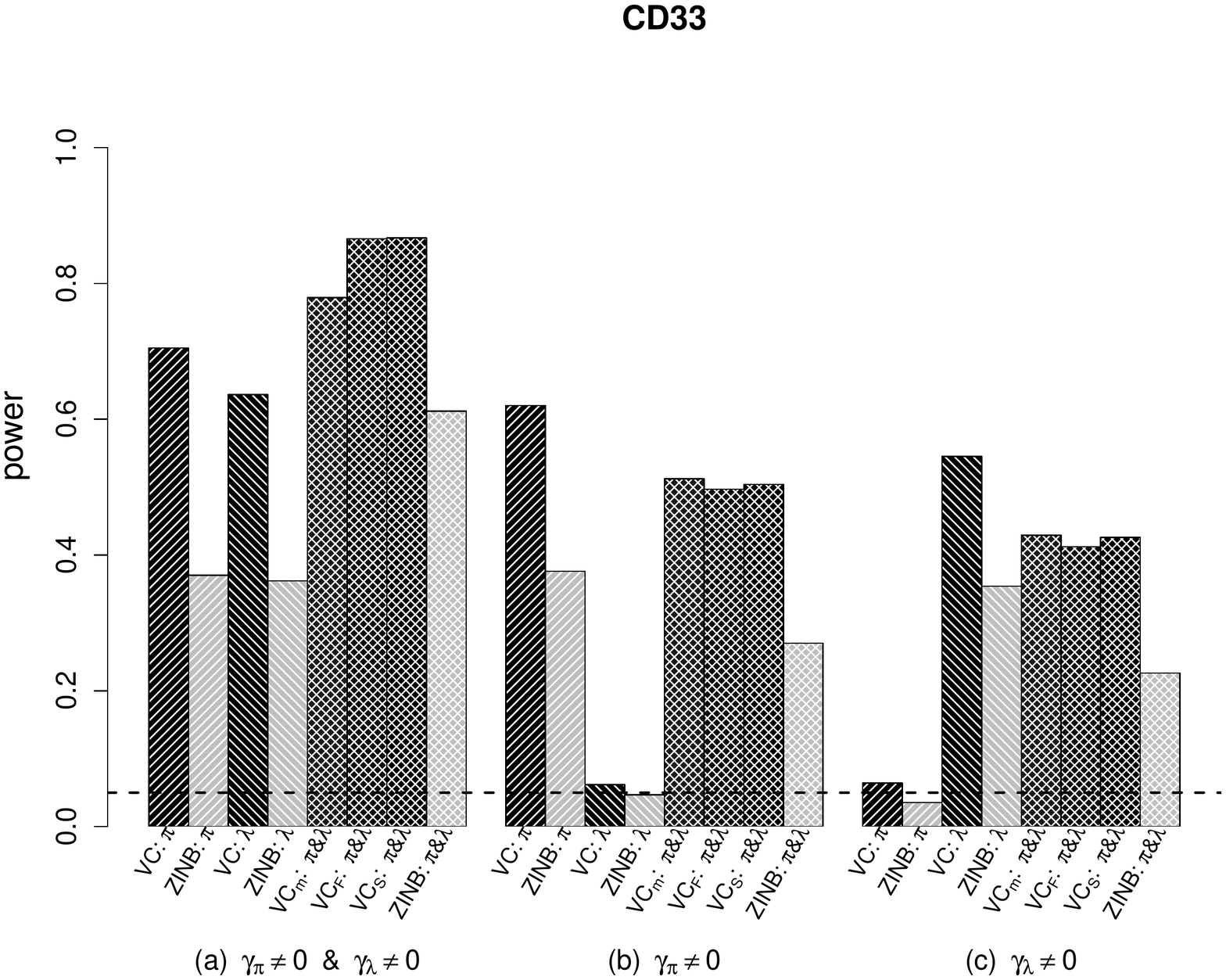}

\caption{Power of different testing procedures when $\mathbf{{G}}$ affects
(a) both mixing proportion and Poisson rate; (b) mixing proportion
only; and (c) Poisson rate only.\label{fig:Power-of-different} }
\end{figure}

Figure \ref{fig:Power-of-different} shows empirical power of different
testing procedures when $\mathbf{{G}}$ is associated with (a) both
$\pi$ and $\lambda$; (b) $\pi$ only; and (c) $\lambda$ only. We
find that our method outperforms the competing methods on power in
most scenarios, especially with those where the true signal is through
the mixing proportion $\pi$ alone. There the APOE plots suggests
that in some LD scenarios the power of our method can outperform the
competitors when the true signal is from $\pi$, but underperform
when the signal is from $\lambda$. The combined test has comparable
performance to competing methods in the lower-dimensional APOE case.
However in the higher dimensional CD33 scenario we find that the power
of our method is consistently superior.

\subsection{Data Application}

\subsubsection{Data Set}

The ROSMAP study is a longitudinal, epidemiologic clinical-pathologic
cohort study following subjects who are free of cognitive impairment
at baseline. It focuses on cognition, tested annually, while also
investigating correlated pathology in postmortem brain tissue analysis.
Further details of the study can be found in Bennett et al. \citep{bennett2012}.
Genotype information was generated using the Affymetrix GeneChip 6.0
genotype platform, as described in \citet{Chibnik2011}. Genotyping
in the original study was limited to self-identified non-Hispanic
Caucasians to reduce population heterogeneity. Remaining population
heterogeneity was controlled by using the first three eigenvectors
of an eigendecomposition of the genotypes. Genotypes were imputed
using BEAGLE software, version 3.3.2 and 1000 Genomes Project Consortium
interim phase I haplotypes, 2010-2011 data freeze. Association analysis
was performed using the imputed genotype dosages.

Among 983 subjects who had died and had postmortem NP pathology measurements
as well as imputed genotypes as of 2015, we restricted to the 970
subjects for whom the covariate `packyears at baseline', a smoking
status indicator, was recorded. We also include sex and age at death
as covariates. The covariates `packyears at baseline' and `age at
death' are included in the models as linear terms. 

To illustrate the utility of our method we applied the test to several
candidate genes, including ABCA7, APOE, CD33, MAPT and PTPRD, which
have previously been shown to be associated with the AD risk. APOE
on cytoband 19q13.32 \citep{cytob} is a well known gene previously
reported to be associated with Alzheimer's risk \citep{Lambert2013}.
The previously reported GWAS SNPs for APOE were rs429358, and rs7412,
which we used to obtain additional SNPs in LD. The additional genes
we analyzed were ABCA7 (19p13.3 ; rs3752246, rs3764650), CD33 (19q13.41;
rs3865444, rs3826656), MAPT (17q21.31; rs1800547, rs3785883, rs8070723)
and PTPRD (9p23; rs560380, rs3764650). These genes and SNPs have also
been found to be associated with Alzheimer's \citep{Lambert2013,giri2016genes}.
The previously reported GWAS SNPs for CD33 were rs3865444, rs3826656.
To select a SNP set for analysis we started with all BEAGLE imputed
SNPs lying within the boundaries of the gene along with those $\pm10000$
bp upstream and downstream. We restricted to those SNPs which matched
the candidate gene on the flag `ensembl\_gene\_id' in the Ensembl.org
GRCh38.11 data. We further restricted to SNPs with a variance equivalent
to an MAF of 0.01 or above. We also restricted to SNPs that were in
LD, with correlation ranging from 0.4 to 0.95, with SNPs previously
reported to be significantly associated with the AD case-control phenotype
in GWAS. These restrictions yields 8 SNPs for APOE and 22 for CD33.

\begin{figure}
\includegraphics[width=8cm]{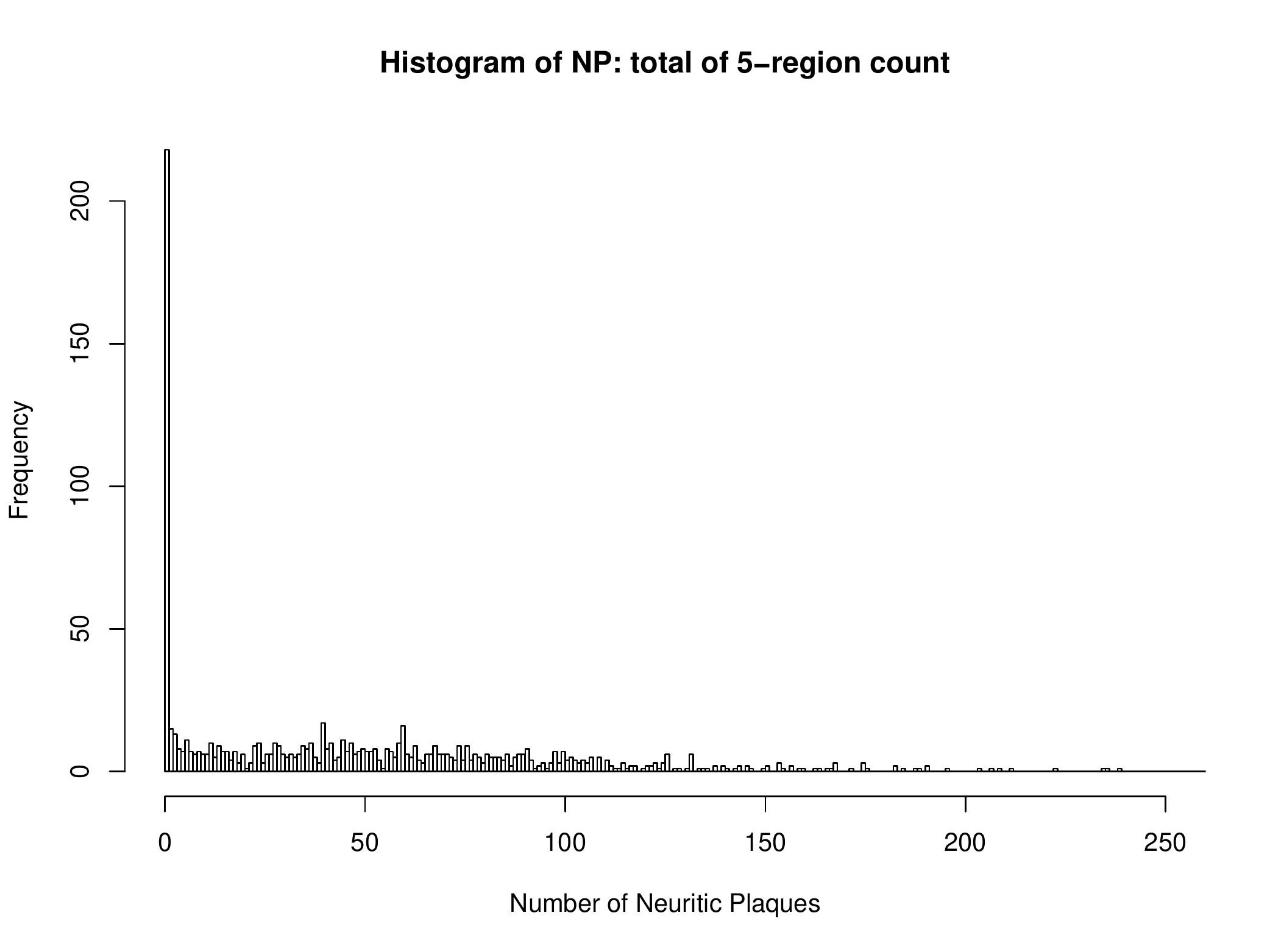}

\caption{Histogram of NP outcome\label{fig:NP-Hist}}
\end{figure}

The phenotype outcome we analyze is a total count measure of neuritic
plaques (NP), taken from five sites: hippocampus CA1, entorhinal cortex,
inferior parietal cortex, midfrontal cortex, and midtemporal cortex.
The site-specific counts are recorded by a technician who observes
tissue pathology slides from each site, and are then summed to obtain
a total count. The total count shows evidence of both zero-inflation
and overdispersion (see figure \ref{fig:NP-Hist}).

\subsubsection{Data Analysis Results}

In Table \ref{tab:DataAnalysis}, we present p-values for these genes
obtained based on our proposed VC test as well as the ZINB Wald test.
Other competing tests were excluded from this analysis due to their
susceptibility to inflated type I error. We take $\alpha=0.05$ as
the nominal size.

\begin{table}
{\scriptsize{}
\begin{tabular}{rrrrrrr}
  \hline
 & \specialcell{$\pi$: \\ VC} & \specialcell{$\pi$: \\ ZINB Wald} & \specialcell{$\lambda$: \\  VC} & \specialcell{$\lambda$: \\ ZINB Wald} & \specialcell{$\pi \& \lambda$: \\ VC std.} & \specialcell{$\pi \& \lambda$: \\  ZINB} \\ 
  \hline
ABCA7 & 2.1E-01 & 2.1E-02 & 1.9E-01 & 6.6E-01 & 1.8E-01 & 1.1E-01 \\ 
  APOE & 1.1E-08 & 2.0E-06 & 1.6E-06 & 1.4E-08 & 5.8E-12 & 4.4E-13 \\ 
  CD33 & 3.5E-01 & 6.7E-01 & 9.9E-01 & 3.9E-01 & 7.4E-01 & 5.8E-01 \\ 
  MAPT & 7.7E-01 & 8.5E-01 & 4.1E-01 & 8.6E-01 & 6.7E-01 & 9.4E-01 \\ 
  PTPRD & 4.0E-01 & 7.6E-01 & 3.0E-04 & 2.7E-01 & 1.6E-03 & 5.2E-01 \\ 
   \hline
\end{tabular}
}{\scriptsize \par}

\smallskip{}

\caption{Data analysis results: p-values of selected genes in ROSMAP data\label{tab:DataAnalysis}}
\end{table}

Looking at the results from our proposed method, significant p-values
for APOE are consistent with past studies which have shown it to be
strongly related to Alzheimers risk. We note that the p-value associated
with $\lambda$ is more extreme than that for $\pi$. Thus the evidence
that APOE is associated with total count of neuritic plaques appears
to be stronger than the evidence APOE is associated with the presence
or absence of plaques. Note however, that at least part of this discrepancy
can be accounted by relative power in the tests for $\pi$ and $\lambda$,
which may be affected both by effect size and by different statistical
considerations for each parameter. For example it is well known that
in tests of binary regression, power is a function of $\pi$, with
lower power as $\pi$ approaches 0 or 1, and the same pattern is expected
to hold here. By contrast, a strong association signal in $\pi$ but
not $\lambda$ might suggest that the biological mechanism is `all-or-nothing',
meaning it may prevent or allow formation of neuritic plaques, but
does not affect the rate of plaque formation if it occurs. However,
we caution against making any conclusions of this sort solely based
on p-values. 

\begin{figure}
\includegraphics[width=10cm]{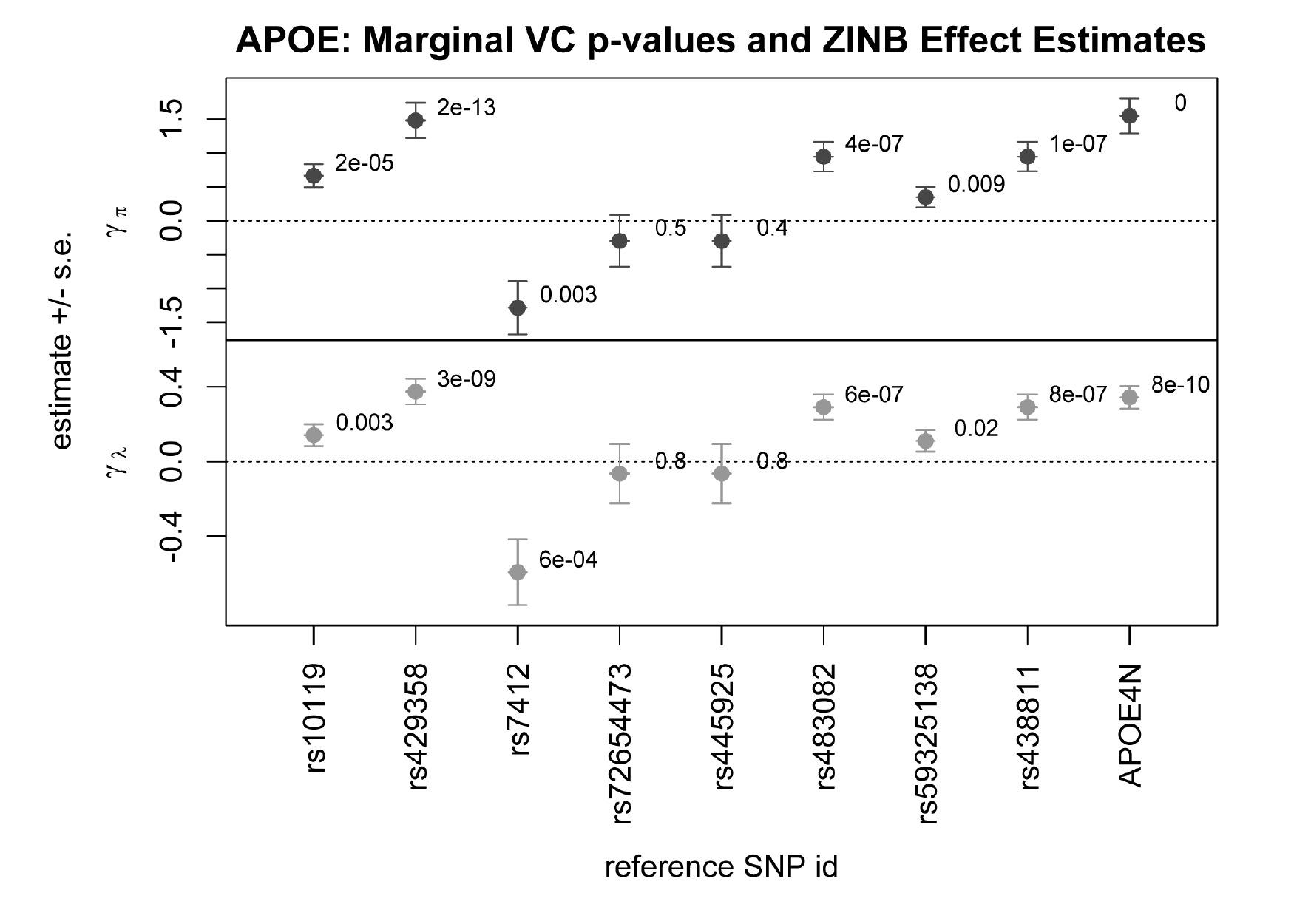}

\caption{Plot of APOE SNP and APOE-$\varepsilon$4 haplotype effects from marginal
models \label{fig:APOE-SNP-eff}}
\end{figure}

To illustrate the breakdown of genetic effects within a SNP set, we
performed ZINB regression for the NP outcome to estimate genetic effects
through $\pi$ and $\lambda$ for each SNP in APOE, using a single-SNP
marginal model approach (but still controlling for the above-mentioned
covariates). We also computed the effect estimate for the regression
against the APOE $\varepsilon4$ allele count, to compare results
for individual SNPs with those for this well-known haplotype. In Figure
\ref{fig:APOE-SNP-eff} we report the maximum likelihood estimates
of genetic effects $\gamma_{\pi}$ and $\gamma_{\lambda}$, along
with their standard errors, and the corresponding $\mbox{{VC}}_{\pi}$
or $\mbox{{VC}}_{\lambda}$ p-value. Results for previously reported
GWAS snps rs429358, and rs7412 show a strong signal in our analysis
in both $\pi$ and $\lambda$. 

Farebrother's method \citep{farebrother1984algorithm}, implemented
in the R statistical software package `CompQuadForm', was used to
calculate the the distribution of $\hat{Q}_{\iota}$ in both the marginal
and the gene-level data application in order to obtain additional
precision for the p-values. Nevertheless, the p-value was calculated
as 0 for the test of $\pi$ for the APOE $\varepsilon4$ allele .

\section{Discussion}

We have derived novel VC tests that can efficiently detect the association
between a set of genetic markers and a zero-inflated count outcome.
We have shown the test to adequately control type I error in many
simulation scenarios and have shown our method to obtain superior
power to competing tests especially with large or sparse sets of markers
with certain LD structures. Additionally we have seen that the proposed
method has feasible application to genomic data from an Alzheimer's
data study. The proposed VC testing procedures appear to adequately
control type I error in all of the correctly specified simulation
settings we investigated. The test is also not sensitive to the presence
of model mis-specification due to overdispersion, which is of practical
importance. This is in part due to the use of perturbation resampling
for ascertaining asymptotic distribution. Even under model mis-specification,
provided that $\mathbf{{X}}$ is independent of $\mathbf{{G}}$, our
proposed procedure can be shown to preserve the type I error. 

The standard GWAS remains one that considers testing individual SNPs
separately. It is possible to take this marginal approach using our
method. However, we believe that the potential to combine SNP effects
at the gene level is likely to increase power in many circumstances.
The linear model proposed in this paper to test the association of
individual genes with a ZIP phenotype is probably adequate for a proportion
of such associations, in particular we detect a signal in the APOE
gene, despite not explicitly coding the APOE $\varepsilon2,$ $\varepsilon3$,
and $\varepsilon4$ haplotypes but instead using additive SNP effects.
However, if it is important to capture interactions or other complex
SNP effects, our method easily admits a generalization which makes
this possible.

With proper choices of $\mathbf{\boldsymbol{{\Psi}}(\cdot)}$, our
proposed procedure corresponds to a kernel machine score test that
can be used to detect non-linear effects. Specifically, suppose $h_{\iota}(\mathbf{{G}})\in\mathcal{{H}}_{K}$
with $\mathcal{{H}}_{K}=\mbox{{span}\ensuremath{\left\{  \psi_{\ell}(\mathbf{{G}\cdot}),\ell=1,...,\mathcal{{J}}\right\} } }$
being a reproducible kernel space generated by a given positive definite
kernel function $K(\cdot,\cdot)$ \citep{cristianini2000introduction}.
Then letting $\boldsymbol{{\Psi}}(\cdot)=[\psi_{\ell}(\cdot),...,\psi_{\mathcal{{J}}}(\cdot)]^{\top}$
leads to a score test of the form $\hat{\mathbf{r}}_{\iota}^{\top}\mathbb{{K}}\hat{\mathbf{r}}_{\iota}$,
where $\hat{\mathbf{r}}_{\iota}=(\hat{{r}}_{\iota,1},...,\hat{{r}}_{\iota,n})^{\top}$and
$\mathbb{{K}}=[K(\mathbf{{G}}_{i},\mathbf{{G}}_{j})]_{n\times n}$
is the observed kernel matrix. Intuitively $K(\cdot,\cdot)$ defines
a measure of genetic similarity between different subjects' genotypes.
Moreover, this transformation projects the original genotype information
via a nonlinear transformation to a larger function space which can
be incorporated into the model linearly. Different choices of kernel
(see \citealp{wu2010}) allow different bases for the non-parametric
function modeling the association between the ZIP outcome and the
$q$ markers in the SNP set, in turn allowing for power to test complex
relationships and interactions.

\pagebreak{}

\section*{Appendix 1: asymptotic distribution}

We next outline the key steps for deriving the asymptotic distribution
for $\sqrt{{n}}\hat{\mathbf{{S}}}\equiv\sqrt{{n}}(\hat{\mathbf{{S}}}_{\pi}^{\top},\hat{\mathbf{{S}}}_{\lambda}^{\top})^{\top}$
under $H_{0}$. To this end, we write $\hat{\mathbf{{S}}}_{\iota}=\widetilde{{\mathbf{S}}}_{\iota}(\widehat{{\boldsymbol{{\beta}}}}_{0})$
and $\bar{\mathbf{{S}}}_{\iota}=\widetilde{{\mathbf{S}}}_{\iota}(\boldsymbol{{\beta}}_{0})=n^{-1}\sum r_{\iota,i}\mathbf{\mathbf{\boldsymbol{\Psi}(}G}_{i})$,
where $\widetilde{{\mathbf{S}}}_{\iota}(\boldsymbol{{\beta}})=n^{-1}\sum r_{\iota}(Y_{i},\mathbf{X}_{i}^{\top}\boldsymbol{{\beta}}_{\pi},\mathbf{X}_{i}^{\top}\boldsymbol{{\beta}}_{\lambda})\mathbf{\mathbf{\boldsymbol{\Psi}(}G}_{i})$,
and $r_{\iota,i}=r_{\iota}(Y_{i},\mathbf{X}_{i}^{\top}\boldsymbol{{\beta}}_{0,\pi},\mathbf{X}_{i}^{\top}\boldsymbol{{\beta}}_{0,\lambda})$
First, by a standard law of large numbers and properties of the maximum
likelihood estimator, we have $\widehat{{\boldsymbol{{\beta}}}}_{0}\to{\boldsymbol{{\beta}}}_{0}$
in probability and $\sqrt{{n}}(\widehat{{\boldsymbol{{\beta}}}}_{0}-{\boldsymbol{{\beta}}}_{0})=n^{-1/2}\sum_{i=1}^{n}\{(\mathbb{\mathbf{\mathbb{{I}}}}_{\pi}^{-1}\mathbf{{U}}_{\pi,i})^{\top},(\mathbb{\mathbf{\mathbb{{I}}}}_{\lambda}^{-1}\mathbf{{U}}_{\lambda,i})^{\top}\}^{\top}+o_{p}(1)$
which converges in distribution to a zero-mean multivariate normal.
On the other hand, it follows from a uniform law of large numbers
\citep{pollard1990} that $\widetilde{{\mathbf{S}}}_{\iota}(\boldsymbol{{\beta}})-\mathbf{{s}}_{\iota}(\boldsymbol{{\beta}})\to0$
in probability uniformly in $\boldsymbol{{\beta}}$, where $\mathbf{{s}}_{\iota}(\boldsymbol{{\beta}})=E\left\{ {r_{\iota}(Y_{i},\mathbf{X}_{i}^{\top}\boldsymbol{{\beta}}_{\pi},\mathbf{X}_{i}^{\top}\boldsymbol{{\beta}}_{\lambda})\mathbf{G}_{i}}\right\} $.
Therefore, $|\hat{\mathbf{{S}}}_{\iota}|=|\hat{\mathbf{{S}}}_{\iota}-\mathbf{{s}_{\iota}}({\boldsymbol{{\beta}}}_{0})|\le\sup_{\boldsymbol{{\beta}}}|\widetilde{{\mathbf{S}}}_{\iota}(\boldsymbol{{\beta}})-\mathbf{{s}_{\iota}(\boldsymbol{{\beta}})}|+|\mathbf{{s}_{\iota}(\widehat{{\boldsymbol{{\beta}}}}_{0})}-\mathbf{{s}}_{\iota}({\boldsymbol{{\beta}}}_{0})|\to0$
in probability. The first equality is due to the fact that $\mathbf{{s}}_{\iota}({\boldsymbol{{\beta}}}_{0})=0$,
which can be verified as follows: 
\begin{eqnarray*}
E\left[r_{\pi,i}\right] & \equiv & E\left[I_{Y_{i}=0}\frac{P(Y_{i}>0)}{P(Y_{i}=0)}(1-\pi_{0,i})-I_{Y_{i}>0}(1-\pi_{0,i})\right]\\
 & = & P(Y_{i}>0)(1-\pi_{0,i})-P(Y_{i}>0)(1-\pi_{0,i})=0\\
E\left[r_{\lambda,i}\right] & \equiv & E\left[I_{Y_{i}=0}\frac{\lambda_{0,i}\pi_{0,i}exp(-\lambda_{0,i})}{P(Y_{i}=0)}-I_{Y_{i}>0}(Y_{i}-\lambda_{0,i})\right]\\
 & = & \left(\frac{\lambda_{0,i}\pi_{0,i}exp(-\lambda_{0,i})}{P(Y_{i}=0)}\right)P(Y_{i}=0)-\left(E[Y_{i}I_{Y_{i}>0}]-\lambda_{0,i}E[I_{Y_{i}>0}]\right)\\
 & = & \lambda_{0,i}\pi_{0,i}e^{-\lambda_{0,i}}-\lambda_{0,i}\pi_{0,i}+\lambda_{0,i}\pi_{0,i}(1-e^{-\lambda_{0,i}})=0.
\end{eqnarray*}

To establish the asymptotic distribution for $\sqrt{{n}}\hat{\mathbf{{S}}}$,
we first note that by the functional central limit theorem \citep{pollard1990},
$\sqrt{{n}}\{\widetilde{{\mathbf{S}}}(\boldsymbol{{\beta}})-\mathbf{{s}}(\boldsymbol{{\beta}})\}$
converges weakly to a zero-mean Gaussian process and thus is equicontinuous
in $\boldsymbol{{\beta}}$. This, together with a Taylor series expansion,
implies that 

\[
\begin{array}{ccl}
\sqrt{n}\hat{\mathbf{S}}_{\iota} & \equiv & \sqrt{n}\left\{ \widetilde{{\mathbf{S}}}_{\iota}(\widehat{{\boldsymbol{{\beta}}}}_{0})-\mathbf{\widetilde{{\mathbf{S}}}_{\iota}({\boldsymbol{{\beta}}}_{0})}\right\} +\sqrt{n}\bar{{\mathbf{{S}}}_{\iota}}\\
 & = & \sqrt{n}(\rho_{\iota,\pi},\rho_{\iota,\lambda})(\widehat{{\boldsymbol{{\beta}}}}_{0}-{\boldsymbol{{\beta}}}_{0})+\sqrt{n}\bar{{\mathbf{{S}}}_{\iota}}\\
 & = & n^{-1/2}\sum_{i}\left\{ \rho_{\iota,\pi}\mathbb{\mathbf{\mathbb{{I}}}}_{\pi}^{-1}\mathbf{{U}}_{\pi,i}+\rho_{\iota,\lambda}\mathbb{\mathbf{\mathbb{{I}}}}_{\lambda}^{-1}\mathbf{{U}}_{\lambda,i}+r_{\iota,i}\mathbf{\mathbf{\boldsymbol{\Psi}(}G}_{i})\right\} +o_{p}(1).
\end{array}
\]
 It then follows from a central limit theorem that$\sqrt{n}\hat{\mathbf{S}}$
converges in distribution to a multivariate normal distribution $MVN(\mathbf{0},\boldsymbol{{\Sigma}})$. 

One can show that if $\mathbf{S}\sim MVN(\mathbf{0},\boldsymbol{\Sigma})$,
then $\mathbf{S}^{\top}\mathbf{S}$ has an asymptotic distribution
as a linear combination of independent $\chi_{1}^{2}$ random variables.
For completeness, we sketch the argument here. First note $\mathbf{S}^{\top}\mathbf{S}=\left(\mathbf{S}^{\top}\boldsymbol{\Sigma}^{-\frac{1}{2}}\right)\boldsymbol{\Sigma}\left(\boldsymbol{\Sigma}^{-\frac{1}{2}}\mathbf{S}\right)$
so we can write $\mathbf{S}^{\top}\mathbf{S}=\mathbf{U}^{\top}\boldsymbol{\Sigma}\mathbf{U}$
where $\mathbf{U}=\boldsymbol{\Sigma}^{-\frac{1}{2}}\mathbf{S}\sim MVN(\mathbf{0},I_{p\times p})$.
Then find the spectral decomposition $\boldsymbol{\Sigma}=\mathbf{A}^{\top}\boldsymbol{\Lambda}\mathbf{A}$,
where $\boldsymbol{\Lambda}$ is diagonal and $\mathbf{A}^{\top}\mathbf{A}=\mathbf{A}\mathbf{A}^{\top}=I_{p\times p}$.
Hence for $\mathbf{V}=\mathbf{A}\mathbf{U}$, $Var[\mathbf{V}]=\mathbf{A}Var[\mathbf{U}]\mathbf{A}^{\top}=\mathbf{A}\mathbf{A}^{\top}=I$.
So $\mathbf{V}$ is multivariate normal with identity covariance,
and the $k^{th}$ element of $\mathbf{V}$, $V_{k}$ is standard normal.
So we have $\mathbf{S}^{\top}\mathbf{S}=\mathbf{V}^{\top}\Lambda\mathbf{V}=\sum_{k}\lambda_{k}V_{k}\cdot V_{k}=\sum_{k}\lambda_{k}\chi_{1}^{2}$.

\pagebreak{}

\section*{Appendix 2: simulation settings}

Specification of $\mathbf{X}_{i}=(X_{1i},X_{2i},X_{3i},X_{4i},X_{5i})^{\top}$:

$X_{1i}\sim binomial(k=2,p=0.5)$

$W_{2i}\sim binomial(2,0.5)$, $Z_{2i}\sim normal(\mu=0,\sigma^{2}=.25)$

$X_{2i}=0.5(W_{2})+Z_{2}$

$Z_{3i}\sim normal(\mu=0,\sigma^{2}=.25))$

$X_{3i}\sim0.1X_{1i}X_{2i}+Z_{3i}$

$W_{4i}\sim binomial(2,0.4)$, $Z_{4i}\sim normal(\mu=0,\sigma^{2}=.25)$ 

$X_{4i}=0.1(W_{4i})-0.2(X_{2i})+0.2(X_{3i})+Z_{4i}$

$W_{5i}\sim0.5(binomial(n,2,0.4)$, $Z_{5i}\sim normal(\mu=0,\sigma^{2}=.25)$

$X_{5}=0.5(W_{5i})+0.15(X_{2i})+0.2(X_{3i})+Z_{5i}$

Genotype-Covariate Dependence Matrix:

$A=\left[\begin{array}{ccccc}
0.08 & 0.5 & 0.0 & 0.5 & 0.8\\
0.09 & 0.4 & 0.1 & 0.0 & 0.0\\
0.0 & 0.0 & 0.3 & 0.6 & 0.9
\end{array}\right]$

\section*{Appendix 3: table of results}

\begin{center}

\resizebox*{!}{\textwidth}{%
{\doublespacing
\begin{sideways}

{\scriptsize

\begin{tabular}{rlrrrrrrrrrrrr}
  \hline
 & Simulation Setting & \specialcell{$\pi$: \\ VC} & \specialcell{$\pi$: ZIP \\  Wald} & \specialcell{$\pi$: ZINB \\Wald} & \specialcell{$\lambda$: \\  VC} & \specialcell{$\lambda$: ZIP \\ Wald} & \specialcell{$\lambda$: ZINB \\ Wald} & \specialcell{$\pi \& \lambda$: \\ VC min-p} & \specialcell{$\pi \& \lambda$: \\ VC Fisher's} & \specialcell{$\pi \& \lambda$: \\ VC std. } & \specialcell{$\lambda_{Poisson}$: \\  Wald HW} & \specialcell{$\pi \& \lambda$: \\  ZIP} & \specialcell{$\pi \& \lambda$: \\  ZINB} \\ 
  \hline
1 & 1a: APOE with X & 0.055 & 0.040 & 0.040 & 0.054 & 0.050 & 0.047 & 0.053 & 0.053 & 0.056 & 0.080 & 0.045 & 0.042 \\ 
  2 & 1b: APOE without X & 0.050 & 0.035 & 0.034 & 0.056 & 0.053 & 0.050 & 0.056 & 0.056 & 0.055 & 0.071 & 0.040 & 0.038 \\ 
  3 & 2a: APOE N = 1000 & 0.055 & 0.035 & 0.034 & 0.061 & 0.046 & 0.043 & 0.051 & 0.055 & 0.061 & 0.089 & 0.039 & 0.037 \\ 
  4 & 2b: APOE N = 2000 & 0.055 & 0.041 & 0.041 & 0.059 & 0.051 & 0.049 & 0.052 & 0.049 & 0.052 & 0.076 & 0.045 & 0.041 \\ 
  5 & 2c: APOE N = 4000 & 0.052 & 0.047 & 0.047 & 0.056 & 0.049 & 0.046 & 0.050 & 0.051 & 0.054 & 0.082 & 0.052 & 0.050 \\ 
  6 & 3a: APOE low corr. & 0.051 & 0.038 & 0.037 & 0.050 & 0.047 & 0.044 & 0.046 & 0.045 & 0.048 & 0.075 & 0.041 & 0.039 \\ 
  7 & 3b: APOE mod. corr. & 0.055 & 0.035 & 0.034 & 0.061 & 0.046 & 0.043 & 0.051 & 0.055 & 0.061 & 0.089 & 0.039 & 0.037 \\ 
  8 & 3c: APOE high corr. & 0.059 & 0.035 & 0.035 & 0.071 & 0.054 & 0.051 & 0.056 & 0.063 & 0.071 & 0.165 & 0.042 & 0.040 \\ 
  9 & 4a: APOE OD low corr. & 0.056 & 0.040 & 0.036 & 0.056 & 0.391 & 0.061 & 0.048 & 0.053 & 0.060 & 0.094 & 0.280 & 0.045 \\ 
  10 & 4b: APOE OD mod. corr. & 0.058 & 0.040 & 0.035 & 0.067 & 0.487 & 0.064 & 0.053 & 0.064 & 0.070 & 0.111 & 0.376 & 0.050 \\ 
  11 & 4c: APOE OD high corr. & 0.052 & 0.035 & 0.030 & 0.115 & 0.847 & 0.066 & 0.072 & 0.079 & 0.092 & 0.229 & 0.765 & 0.046 \\ 
  12 & \specialcell{5a: APOE Alt. $\pi \& \lambda$} & 0.509 & 0.262 & 0.260 & 0.609 & 0.589 & 0.574 & 0.661 & 0.759 & 0.780 & 0.909 & 0.730 & 0.722 \\ 
  13 & \specialcell{5b: APOE Alt. $\pi$} & 0.390 & 0.218 & 0.216 & 0.058 & 0.051 & 0.048 & 0.290 & 0.295 & 0.303 & 0.261 & 0.159 & 0.153 \\ 
  14 & \specialcell{5c: APOE Alt. $\lambda$} & 0.062 & 0.035 & 0.035 & 0.481 & 0.553 & 0.539 & 0.363 & 0.362 & 0.377 & 0.412 & 0.408 & 0.399 \\ 
  15 & 1a: CD33 with X & 0.052 & 0.033 & 0.032 & 0.057 & 0.055 & 0.052 & 0.051 & 0.052 & 0.053 & 0.114 & 0.038 & 0.036 \\ 
  16 & 1b: CD33 without X & 0.053 & 0.027 & 0.027 & 0.054 & 0.056 & 0.053 & 0.054 & 0.056 & 0.055 & 0.096 & 0.040 & 0.037 \\ 
  17 & 2a: CD33 N = 1000 & 0.054 & 0.029 & 0.028 & 0.058 & 0.048 & 0.045 & 0.061 & 0.057 & 0.060 & 0.149 & 0.030 & 0.028 \\ 
  18 & 2b: CD33 N = 2000 & 0.056 & 0.039 & 0.039 & 0.051 & 0.050 & 0.047 & 0.053 & 0.051 & 0.052 & 0.132 & 0.041 & 0.039 \\ 
  19 & 2c: CD33 N = 4000 & 0.048 & 0.053 & 0.053 & 0.048 & 0.044 & 0.042 & 0.047 & 0.049 & 0.048 & 0.158 & 0.045 & 0.043 \\ 
  20 & 3a: CD33 corr. low & 0.055 & 0.030 & 0.030 & 0.051 & 0.051 & 0.048 & 0.050 & 0.052 & 0.054 & 0.119 & 0.037 & 0.037 \\ 
  21 & 3b: CD33 corr. mod. & 0.054 & 0.029 & 0.028 & 0.058 & 0.048 & 0.045 & 0.061 & 0.057 & 0.060 & 0.149 & 0.030 & 0.028 \\ 
  22 & 3c: CD33 corr. high & 0.057 & 0.025 & 0.025 & 0.065 & 0.051 & 0.049 & 0.061 & 0.057 & 0.060 & 0.339 & 0.035 & 0.034 \\ 
  23 & 4a: CD33 OD corr. low & 0.050 & 0.035 & 0.032 & 0.063 & 0.417 & 0.061 & 0.056 & 0.056 & 0.060 & 0.130 & 0.281 & 0.047 \\ 
  24 & 4b: CD33 OD corr. mod. & 0.053 & 0.031 & 0.028 & 0.059 & 0.593 & 0.061 & 0.053 & 0.060 & 0.058 & 0.173 & 0.450 & 0.037 \\ 
  25 & 4c: CD33 OD corr. high & 0.049 & 0.030 & 0.029 & 0.065 & 0.920 & 0.062 & 0.060 & 0.060 & 0.064 & 0.323 & 0.847 & 0.043 \\ 
  26 & \specialcell{5a: CD33 Alt. $\pi \& \lambda$} & 0.705 & 0.370 & 0.370 & 0.636 & 0.372 & 0.362 & 0.779 & 0.866 & 0.867 & 0.747 & 0.621 & 0.612 \\ 
  27 & \specialcell{5b: CD33 Alt. $\pi$} & 0.620 & 0.377 & 0.376 & 0.062 & 0.050 & 0.047 & 0.512 & 0.497 & 0.504 & 0.249 & 0.276 & 0.270 \\ 
  28 & \specialcell{5c: CD33 Alt. $\lambda$} & 0.064 & 0.035 & 0.035 & 0.545 & 0.366 & 0.354 & 0.429 & 0.412 & 0.426 & 0.300 & 0.234 & 0.226 \\ 
   \hline
\end{tabular}
}

\end{sideways}
}
}
\par\end{center}

{\scriptsize \par}

\textbf{Table 1:} Table of p-value results. See text for detailed
descriptions of simulation scenarios.

\pagebreak{}

\section*{Supplement 1: gene SNP sets}

Gene sets for simulations and data analysis:

ABCA7 {[}46 SNPs{]}

rs930232, rs930231, rs3848640, rs3764642, rs4147904, rs3764645, rs4622634,
rs4147909, rs3752237, rs4147912, rs4147914, rs7408475, rs12151021,
rs3764651, rs3764652, rs3829687, rs3752242, rs3752243, rs3745842,
rs11671895, rs881768, rs3752246, rs2279796, rs2868065, rs10411696,
rs11671157, rs4147922, rs4147923, rs1968456, rs4147924, rs558820627,
rs4807499, rs4147930, rs4147931, rs4147932, rs1609436, rs4147934,
rs4147936, rs2242437, rs2242436, rs4147938, rs905149, rs12981369,
rs1610096, rs3764653, rs1801284

APOE {[}8 SNPs{]} 

rs10119, rs429358, rs7412, rs72654473, rs445925, rs483082, rs59325138,
rs438811

CD33 {[}22 SNPs{]} 

rs273637, rs273638, rs273639, rs273640, rs1399837, rs3826656, rs1710398,
rs1697553, rs3865444, rs2459141, rs2455069, rs7245846, rs34813869,
rs1354106, rs35112940, rs10409348, rs273652, rs1697531, rs169275,
rs273649, rs273648, rs273646

MAPT {[}62 SNPs{]} 

rs7210728, rs4792891, rs9303523, rs28646281, rs1560312, rs3785879,
rs9915721, rs9899833, rs3785880, rs8080903, rs1560313, rs9904290,
rs1001945, rs8078967, rs2435205, rs242557, rs242562, rs878918, rs242554,
rs2664006, rs9896485, rs10514889, rs1800547, rs754593, rs6503453,
rs713522, rs2471737, rs2435200, rs2435201, rs8067056, rs12946693,
rs56087321, rs41543317, rs62062273, rs916896, rs7521, rs7220988, rs8070723,
rs3744456, rs66660193, rs71375325, rs3785883, rs2435206, rs2435207,
rs7209707, rs63750072, rs2435209, rs2435210, rs2435211, rs11568305,
rs7216893, rs2435202, rs73317025, rs73317026, rs11079728, rs41543512,
rs60969130, rs66499584, rs67676322, rs16940797, rs11652638, rs1078997

PTPRD {[}62 SNPs{]} 

rs408960, rs324512, rs324513, rs324514, rs192973, rs635725, rs493588,
rs634921, rs634098, rs580780, rs324481, rs324482, rs324483, rs324484,
rs324485, rs182720, rs324486, rs324487, rs324488, rs324474, rs324475,
rs324476, rs324478, rs324479, rs324480, rs526677, rs526675, rs324473,
rs324472, rs324470, rs324469, rs324468, rs484454, rs674362, rs369166,
rs523872, rs500091, rs2784543, rs2777489, rs3004230, rs828405, rs448616,
rs557369, rs324465, rs172863, rs324466, rs324467, rs472324, rs324543,
rs324542, rs10759040, rs324546, rs324545, rs324544, rs560380, rs324541,
rs324540, rs1373806, rs324548, rs17666445, rs2007138, rs436929

\newpage{}\textbf{8 imputed SNPs in APOE:} 

\{rs10119, \textbf{rs429358}, \textbf{rs7412}, 19-50106239, rs445925,
rs483082, rs59325138, \textbf{rs438811}\}

(3 SNPs, $G_{Yi}$ (in bold), were associated with the outcome in
the simulated data.)

\begin{figure}[!h]
\includegraphics[width=8cm]{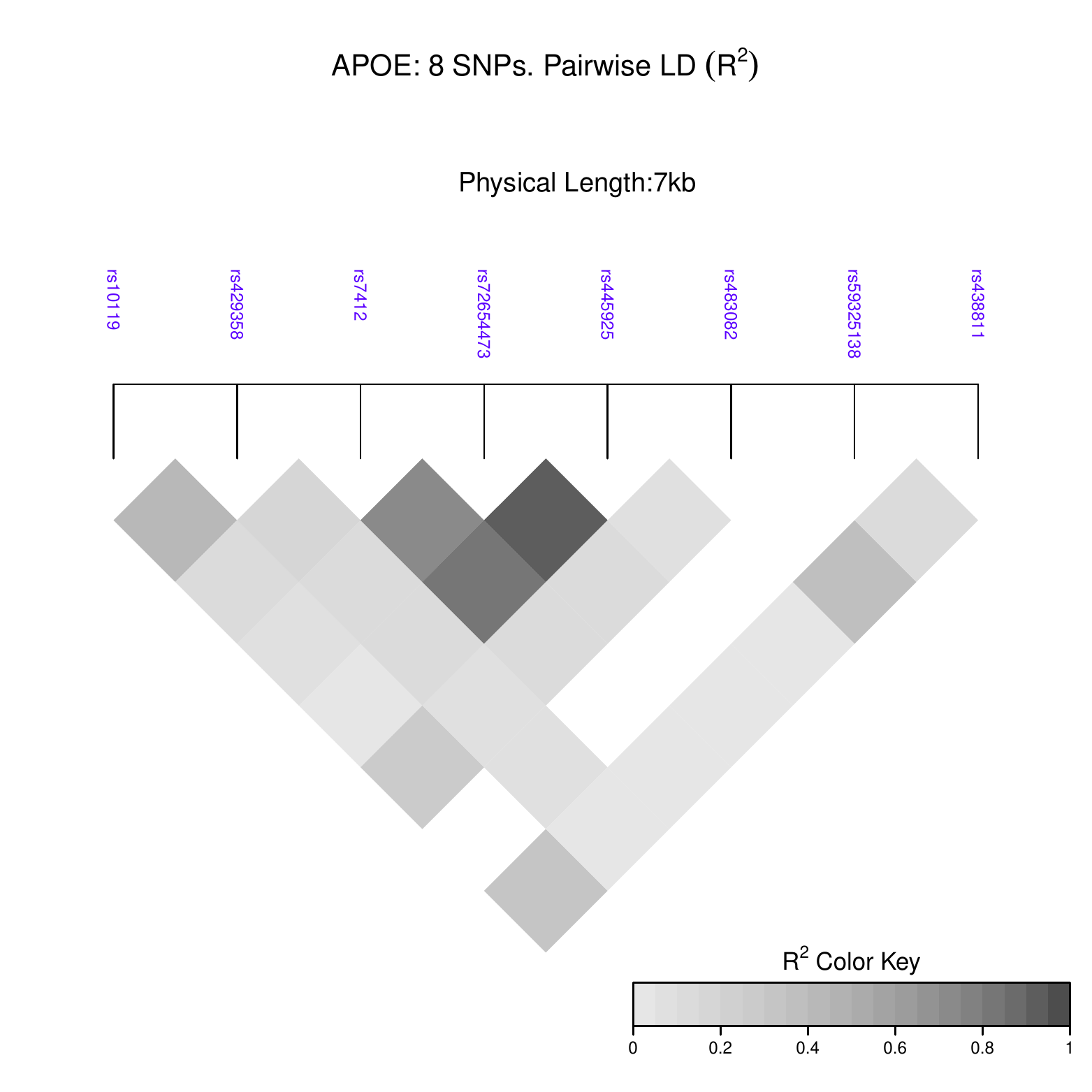}

\caption{LD structure of selected APOE SNPs: HAPGEN Data}
\end{figure}

\begin{figure}[!h]
\includegraphics[width=8cm]{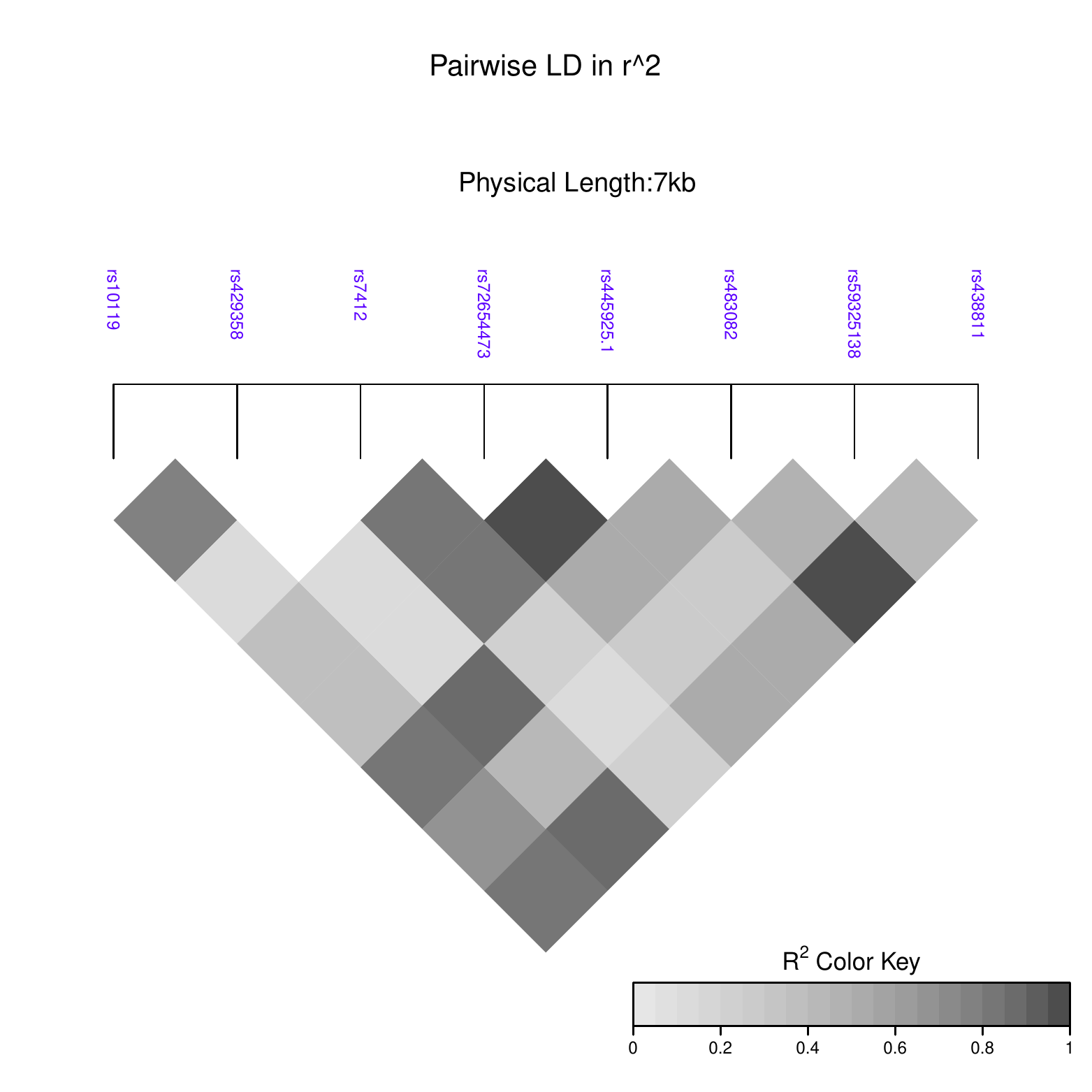}

\caption{LD structure of selected APOE SNPs: ROSMAP Data}
\end{figure}

\newpage{}\textbf{22 imputed SNPs in CD33: }

\{rs273637, rs273638, rs273639, \textbf{rs273640}, rs1399837, rs3826656,
rs1710398, rs1697553, 19-56419774, rs2459141, 19-56420453, rs7245846,
\textbf{rs34813869}, rs1354106, rs35112940, \textbf{rs10409348}, rs273652,
rs1697531, rs169275, rs273649, rs273648, rs273646\}

(3 SNPs, $G_{Yi}$ (in bold), were associated with the outcome in
the simulated data.)

\begin{figure}[H]
\includegraphics[width=8cm]{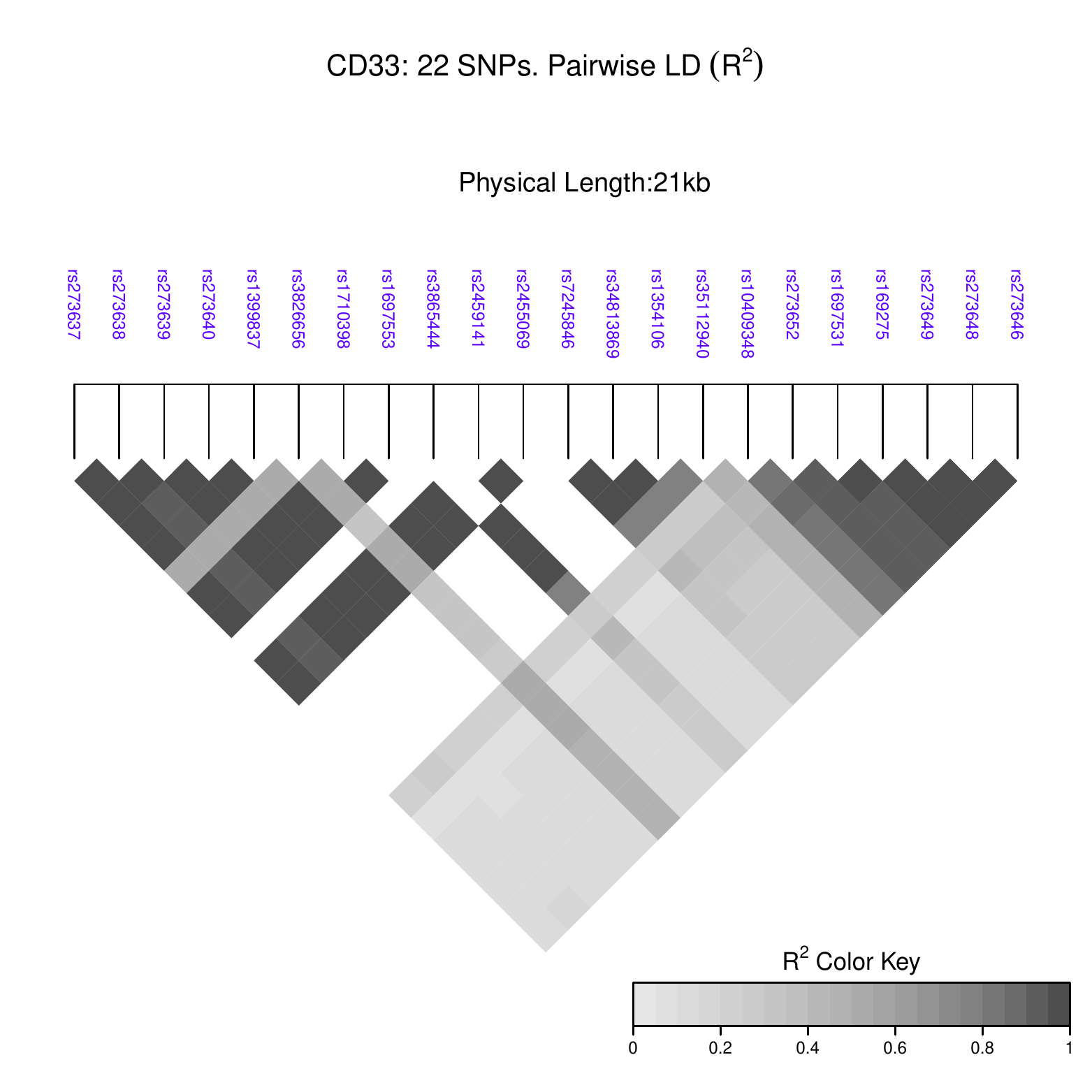}

\caption{LD structure of selected CD33: HAPGEN data}
\end{figure}

\begin{figure}[H]
\includegraphics[width=8cm]{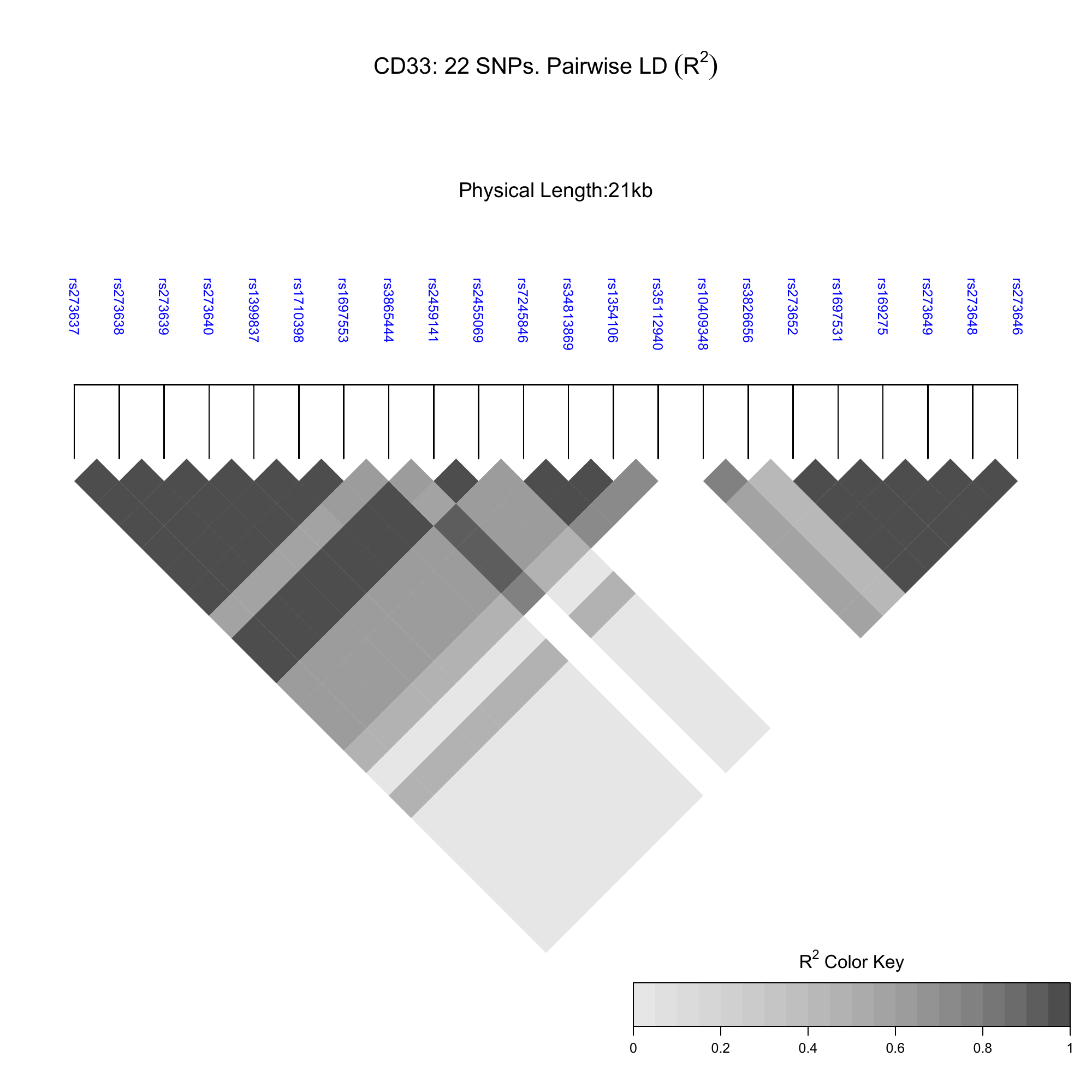}

\caption{LD structure of selected CD33 SNPs: ROSMAP Data}
\end{figure}

\newpage{}

\bibliographystyle{apalike}

\end{document}